\documentclass[11pt,a4paper]{article}
\usepackage[utf8]{inputenc}
\usepackage[english]{babel}
\usepackage[width=\textwidth,labelfont=bf]{caption}
\usepackage[top=30mm, bottom=30mm, left=30mm, right=30mm]{geometry}
\usepackage{setspace,color,enumitem,amsmath,amsfonts,amssymb,amsthm,authblk}
\usepackage{graphicx,hyperref,doi,siunitx,booktabs,threeparttable,afterpage}

\usepackage{algorithm}
\usepackage{algpseudocode}
\usepackage{comment}
\usepackage{orcidlink} 

\usepackage[square,numbers]{natbib}
\hypersetup{
  colorlinks,breaklinks,
  linkcolor={blue},
  urlcolor={blue},
  anchorcolor={blue},
  citecolor={blue}
  }

\newcommand{\T}{\top}

\def\sym#1{\ifmmode^{#1}\else\(^{#1}\)\fi}

\begin{document}

\title{\textbf{%
Confidence-weighted integration of human and machine judgments for superior decision-making
}}

\author[1,*,\orcidlink{0000-0002-9598-0616}]{Felipe Y\'{a}\~{n}ez}
\author[2,\orcidlink{0000-0002-5297-2114}]{Xiaoliang Luo}
\author[1,\orcidlink{0009-0000-3456-6190}]{Omar Valerio Minero}
\author[3,\orcidlink{0000-0002-7883-7076}]{Bradley C. Love}

\affil[1]{Max Planck Institute for Neurobiology of Behavior -- caesar, 
Bonn, Germany}
\affil[2]{Department of Experimental Psychology, University College London, 
London, United Kingdom}
\affil[3]{Los Alamos National Laboratory, Los Alamons, NM, United States}
\affil[*]{%
Correspondence:~\href{email:felipe.yanez@mpinb.mpg.de}{felipe.yanez@mpinb.mpg.de}
}

\date{}
\maketitle

\begin{abstract}
\noindent
Large language models (LLMs) can surpass humans in certain forecasting tasks. What role does this leave for humans in the overall decision process? One possibility is that humans, despite performing worse than LLMs, can still add value when teamed with them. A human and machine team can surpass each individual teammate when team members’ confidence is well-calibrated and team members diverge in which tasks they find difficult (i.e., calibration and diversity are needed). We simplified and extended a Bayesian approach to combining judgments using a logistic regression framework that integrates confidence-weighted judgments for any number of team members. Using this straightforward method, we demonstrated its effectiveness in both image classification and neuroscience forecasting tasks. Combining human judgments with one or more machines consistently improved overall team performance. Our hope is that this simple and effective strategy for integrating the judgments of humans and machines will lead to productive collaborations.
\end{abstract}

\subsection*{Keywords}
Human-AI Collaboration, 
Teaming,
Complementarity,
Large Language Models (LLMs), 
Object Recognition,
Neuroscience,
Decision Making

\section{Introduction}
Modern environments increasingly stretch our ability to process the vast amounts of information available to us~\citep{eppler2004_overload,bawden2009_overload}.
In contrast, machine systems can often take advantage of vast information resources~\citep{deeplearning2015,
alphago,alphafold,news-summarization}.
As machines reach superhuman 
performance levels~\citep{alphafold,gpt2,LLMs_few-shots},
one concern is whether machines will supplant 
human judgment in critical areas~\citep{brynjolfsson2014,FREY2017254}.

One potential solution is forming human--machine teams in which judgments from humans and machines are integrated~\citep{Vaccaro2024,hemmer2024complementarity,steyvers2022hai}. 
It might be possible that humans can contribute to and make the overall team better even when their performance is worse on average than their machine teammates.

We will begin evaluating this possibility in an object recognition task where human and
machine performance overlap according to experimental conditions~\citep{steyvers2022hai}.
Human--machine teaming combines the individual judgments of humans and machines.
We will then evaluate a knowledge-intensive task in which large language models (LLMs) 
surpass humans in predicting the outcomes of neuroscience studies~\citep{luo_large_2024},
posing a real challenge for effective team collaboration.
{\em Complementarity} is realized when a team's performance improves beyond that of 
either teammate alone~\citep{hemmer2024complementarity,steyvers2022hai}.
We investigate whether human--LLM teams outperform LLMs 
even when humans have inferior performance compared to LLMs.
There are two key conditions for team complementarity to be fulfilled~\citep{
steyvers2022hai,luo_large_2024,steyvers2024calibrationgapmodelhuman}.
The first requirement is {\em calibration} of confidence.
This implies that when humans and LLMs have a 
higher degree of confidence in their judgments, 
the accuracy of those judgments tends to be greater~\citep{luo_large_2024}.
The second requirement is classification {\em diversity} among team members.
Diversity holds when the errors in classification made by humans and LLMs are 
not the same~\citep{luo_large_2024}.

Previous work~\citep{steyvers2022hai} has explored the conditions for
complementarity in the context of object recognition.
Humans outperformed machines in the classification of natural images with low levels of noise, 
raising the question of whether a combined approach could achieve 
superhuman performance.
They developed a Bayesian model that integrates 
the judgments of humans and machines.
With this approach, human--machine complementarity was observed.
However, the combination model is computationally expensive and 
challenging to extend to additional teammates.
Ideally, a model that combines the judgments of humans and machines
should be adaptable and scalable, easily interpretable, 
and allow for any number of teammates.

Here, we aim to offer this ideal solution to human--machine teaming 
while evaluating complementarity in an object recognition task, 
and a knowledge-intensive task that is not based on perceptual judgment. 
Critically, in both scenarios, humans were surpassed by machine systems.
Foreshadowing our results, we find support for effective human--machine teaming through our 
resource-efficient procedure.
Our procedure comprises a logistic-regression-based strategy that provides
confidence-weighted integration of teammates' predictions 
for any number of team members.
Our approach is particularly well-suited for combining human and 
machine judgments and assessing their contribution in predictive tasks.
\section{Results}
In this study, we explore whether humans can contribute to decisions when 
machine models, such as LLMs, are superior to them.  
We developed a logistic-regression-based method that integrates a weighted average
of judgments from teammates, whether humans or machines.
The proposed approach adheres to similar previously reported principles~\citep{steyvers2022hai}, 
offering a number of advantages: it is easy to use, flexible, and resource-efficient
(details can be found in the \nameref{sec:methods} section).
We evaluate our method in an object recognition task where effective
human--machine collaboration has been demonstrated~\citep{steyvers2022hai}.
We then shift our focus to BrainBench~\citep{luo_large_2024} because LLMs
significantly outperform human experts, 
posing a real challenge for team collaboration.

\begin{figure*}[ht!]
\centering\includegraphics[width=0.9\textwidth]{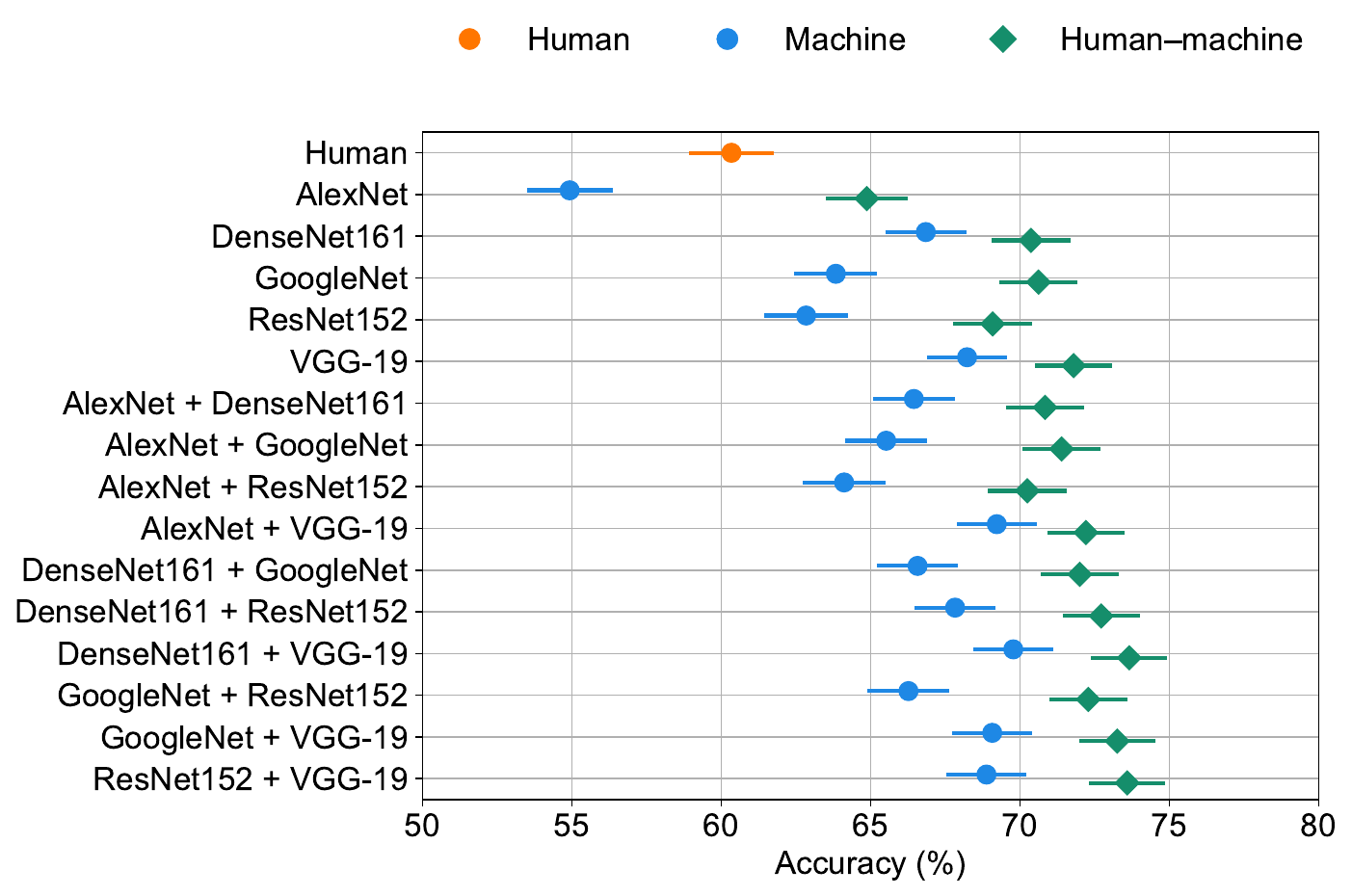}
\caption{\textbf{Performance of the confidence-weighted logistic
combination model in the noisy object recognition task~\protect\citep{steyvers2022hai}.}
\newline
Accuracy results on high levels of image noise ($\Omega=125$) 
with the logistic combination model. 
Human--machine teams (green points) consistently outperform teams without humans (blue points). 
Each data point corresponds to the average across 7\,239 image evaluations. 
Error bars represent standard error of the mean using a binomial model.}
\label{fig:imagenet_125_logistic_predictions_base_experiment}
\end{figure*}

\subsection*{Performance is improved when a human collaborates in a machine-only team}
We first assessed the performance of human--machine teams in the classification 
of noisy natural images~\citep{steyvers2022hai}.
The images were distorted by phase noise at each spatial frequency, 
where the phase noise was uniformly distributed in the interval 
$[-\Omega,\Omega]$~\citep{NEURIPS2018_0937fb58}.
We considered two noise levels: images distorted by low ($\Omega=80$) and 
high ($\Omega=125$) noise.
In the case of low noise, machines are surpassed by humans 
($t(4)=-4.77$, $P<0.01$).
The Bayesian combination model~\citep{steyvers2022hai} demonstrated human--machine complementarity  
(Figure~{\ref{fig_supp:imagenet_80_reproducing_original_study}}).
Despite its simpler setup, our confidence-weighted logistic combination model 
was able to provide team complementarity not only for human--machine teams 
but also for machine--machine teams
(Figure~{\ref{fig_supp:imagenet_80_logistic_predictions_base_experiment}}).
Furthermore, our approach outperformed the Bayesian combination model 
(Welch's $t(20.87)=2.91$, $P<0.01$). 

Figure~{\ref{fig:imagenet_125_logistic_predictions_base_experiment}} shows
the performance of our confidence-weighted logistic combination model
in the case of high noise, where most machines outperform humans
(Figure~{\ref{fig_supp:imagenet_calibration}}).
Of primary interest was whether teams including humans performed better than 
machine-only teams. 
We assessed machine-only teams comprising either one or two machines.
Adding a human teammate to machine-only teams always improved 
the team's performance (Welch's $t(22.78)=4.70$, $P<0.0001$).

\begin{figure*}[ht!]
\centering\includegraphics[width=0.85\textwidth]{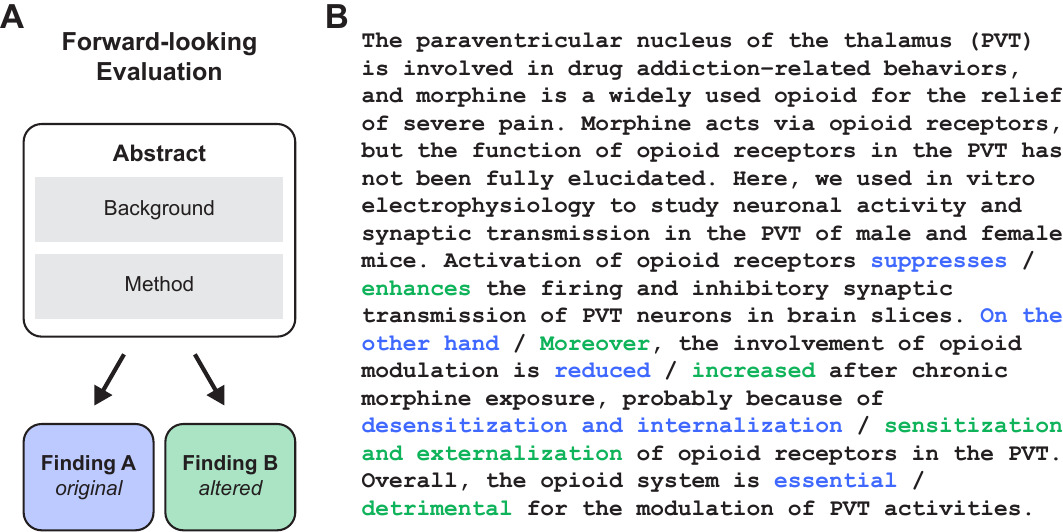}
\caption{\textbf{Assessing Humans and LLMs using BrainBench~\protect\citep{luo_large_2024}.}
\newline
(A) The benchmark comprises test cases constructed from the \emph{Journal of Neuroscience} abstracts. Abstracts consist of background, methods, and results. The test-taker chose which of two versions of the abstract was the original version. The altered version maintained coherency while significantly altering the results. The 100 test cases considered here were constructed by GPT-4 with human oversight and quality control.
\newline
(B) An example test case. Humans were instructed to select which version of the abstract was the original by clicking on either blue or green text to select that set of options. Test cases varied in the numbers of alternatives, but a single click will choose all options of the same color. After their choice, humans indicated their confidence. LLMs chose the version of the abstract that had the lower perplexity score and their confidence was assessed by the absolute difference in perplexity of the two options.}
\label{fig:brainbench}
\end{figure*}

\begin{figure*}[ht!]
\centering\includegraphics[width=\textwidth]{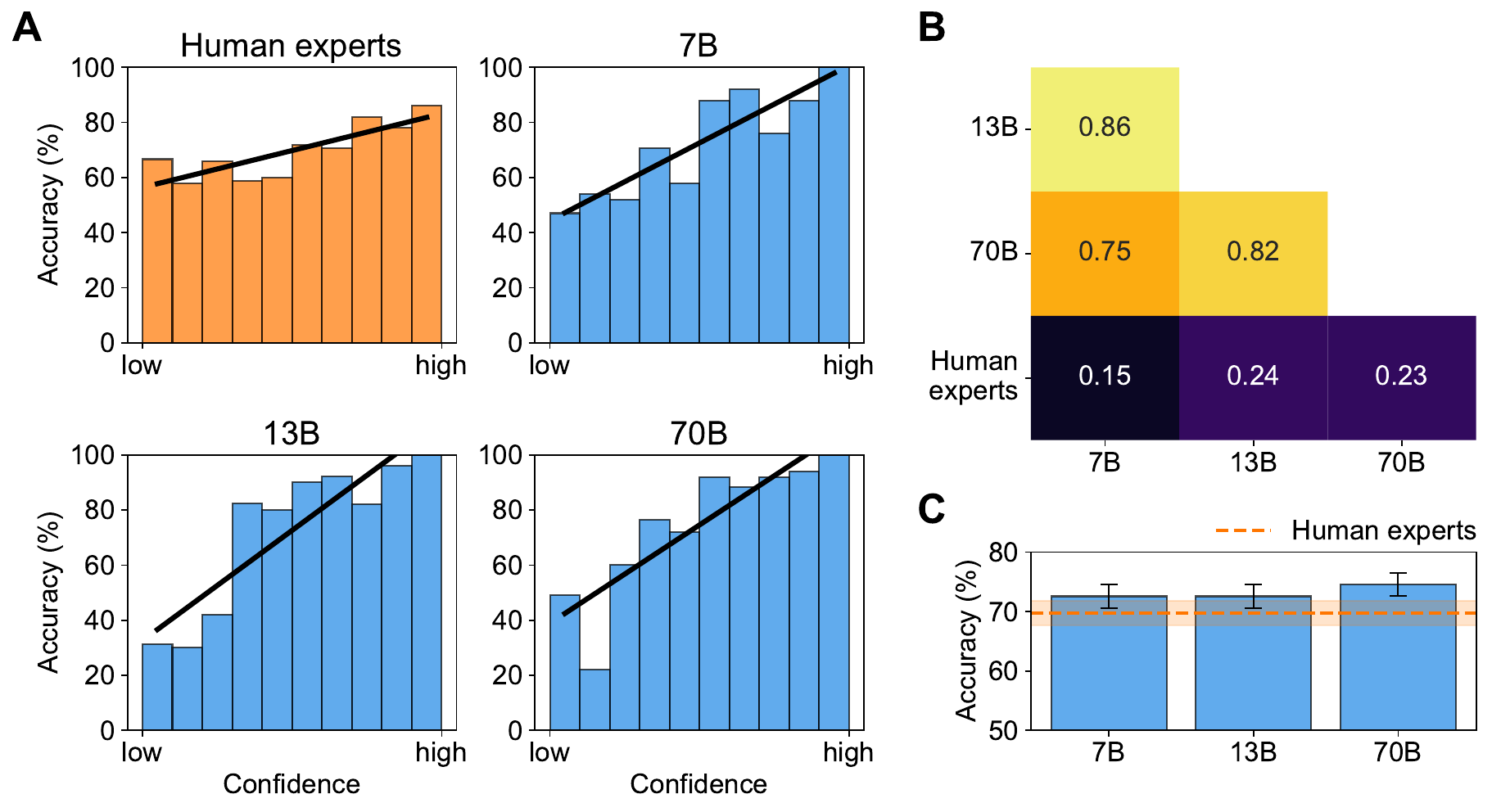}
\caption{\textbf{Conditions for effective collaboration between human experts and LLMs were satisfied.}
\newline
(A) When human experts and LLMs were confident in their BrainBench judgments, they were more likely to be correct. Confidence ratings were sorted into equal bins, and the mean accuracy for each bin was plotted. The positive slope of the black regression lines for humans and Llama2 chat models (7B, 13B, and 70B) indicates well-calibrated confidence~\protect\citep{luo_large_2024,KEREN1991217,Baranski1994,tian-etal-2023-just}, meaning higher confidence correlates with higher accuracy.
\newline
(B) Item difficulty Spearman correlations among LLMs and human experts. For LLMs, difference in perplexity between incorrect and correct abstracts was used to determine the relative difficulty of test cases. Mean accuracy was used for human experts. LLMs align more with each other than humans, which implies human--machine teams will be diverse. Heatmap color scale ranges from 0.1 to 0.9.
\newline
(C) LLMs surpass human experts on BrainBench overall. Error bars represent standard error of the mean using a binomial model.}
\label{fig:brainbench_results_llama2_chat}
\end{figure*}

Bayesian integration is marked by combining judgments based on confidence ratings from team members, 
whether human or machine.
Our confidence model can use this information as well.
However, it remains unclear whether confidence weighting is crucial, or if improved team performance simply reflects an averaging or wisdom-of-the-crowd effect.
This question has not been asked with this dataset, but is straightforward to evaluate using variations of our regression approach.
We found that removing confidence in our approach did not negatively impact team performance in this noisy object recognition task (Figure~{\ref{fig_supp:imagenet_125_logistic_predictions_no_confidence_experiment}}).
For human--machine teams, there is no difference between our standard model and the variant that does not weight by confidence (Welch's $t(26.71)=0.29$, $P=0.77$). 
Surprisingly, for machine--machine teams, the no-confidence model outperformed the standard
(Welch's $t(17.47)=2.58$, $P<0.01$).
Similar performance between signed confidence and sign-only features indicates that confidence does not contribute to improved predictions. This is a consequence of machine classifiers being skewed toward high probability scores (Figure~{\ref{fig_supp:imagenet_confidence_analysis}}).
On the one hand, these results demonstrate the versatility of our approach in that different variations of our model can assess the basis for the success of human--machine teams. 
On the other hand, the fact that the object recognition dataset did not require confidence-weighted integration motivates considering another dataset for which we know that confidence of humans and machines are calibrated in that higher confidence is associated with higher accuracy. The next dataset considered provides a valuable test of our standard confidence-weighted model. 

\begin{figure*}[ht!]
\centering\includegraphics[width=0.75\textwidth]{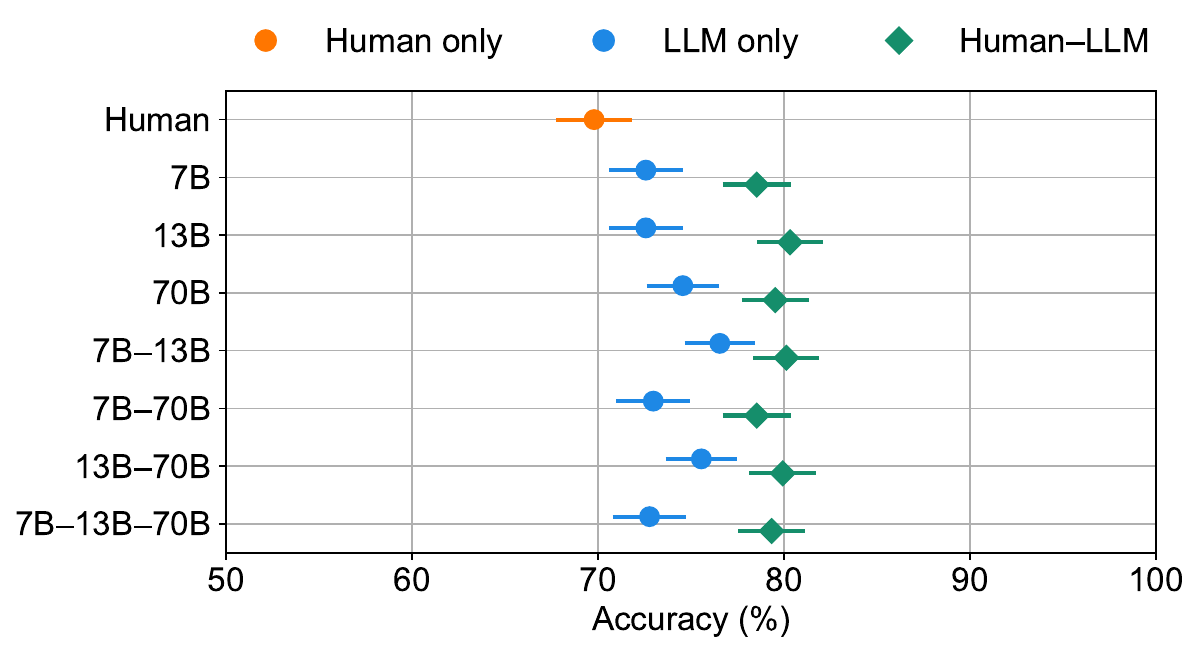}
\caption{\textbf{Performance of all possible teams using the confidence-weighted logistic combination model.}
\newline
Adding a human to a team with one or more machines (blue points) always has a benefit (green points). Llama2 chat 7B, 13B, and 70B models are considered. 
Each data point corresponds to the average across 503 test case evaluations.
Error bars represent standard error of the mean using a binomial model.}
\label{fig:brainbench_logistic_predictions_base_experiment}
\end{figure*}

\subsection*{Calibration of confidence and classification diversity afford reliable human--LLM teaming}
Our contribution also relies on previous efforts that developed 
BrainBench~\citep{luo_large_2024} to assess the capacity of humans and LLMs 
to predict the outcomes of neuroscience studies.
The benchmark includes test cases based on abstracts from
the \emph{Journal of Neuroscience}.
Each test case contains an original abstract and an altered version 
(Figure~{\ref{fig:brainbench}}).
The BrainBench task is to identify the correct study outcome by choosing 
between the original abstract and its altered counterpart.

We evaluated the conditions for effective collaboration (i.e., complementarity), 
namely calibration of confidence and classification diversity among team members.
Both humans and LLMs were calibrated in that accuracy was 
positively correlated with confidence
(Figure~{\ref{fig:brainbench_results_llama2_chat}A}).
Diversity held in that LLMs and humans differed on which test items led to errors
(Figure~{\ref{fig:brainbench_results_llama2_chat}B}).
In terms of accuracy (Figure~{\ref{fig:brainbench_results_llama2_chat}C}), 
LLMs numerically surpassed humans by a small margin ($t(2)=5.20$, $P<0.05$). 
Thus, we can consider whether humans can benefit teams consisting of 
machines that perform comparably or better.

Similarly as in the image classification task, we investigated 
whether teams including humans performed better than LLM-only teams
in the prediction of neuroscience results. 
All 15 possible team combinations, 
ranging from individual teammates to a 4-way human--LLM team,
were considered (Figure~{\ref{fig:brainbench_logistic_predictions_base_experiment}}).
Adding a human teammate to LLM-only teams always improved 
the team's performance (Welch's $t(8.29)=8.24$, $P<0.0001$).
Pairing a human with an LLM led to a more effective (i.e., accurate) team 
than pairing the LLM with a different LLM ($t(2)=10.39$, $P<0.01$).

\begin{figure*}[ht!]
\centering\includegraphics[width=0.75\textwidth]{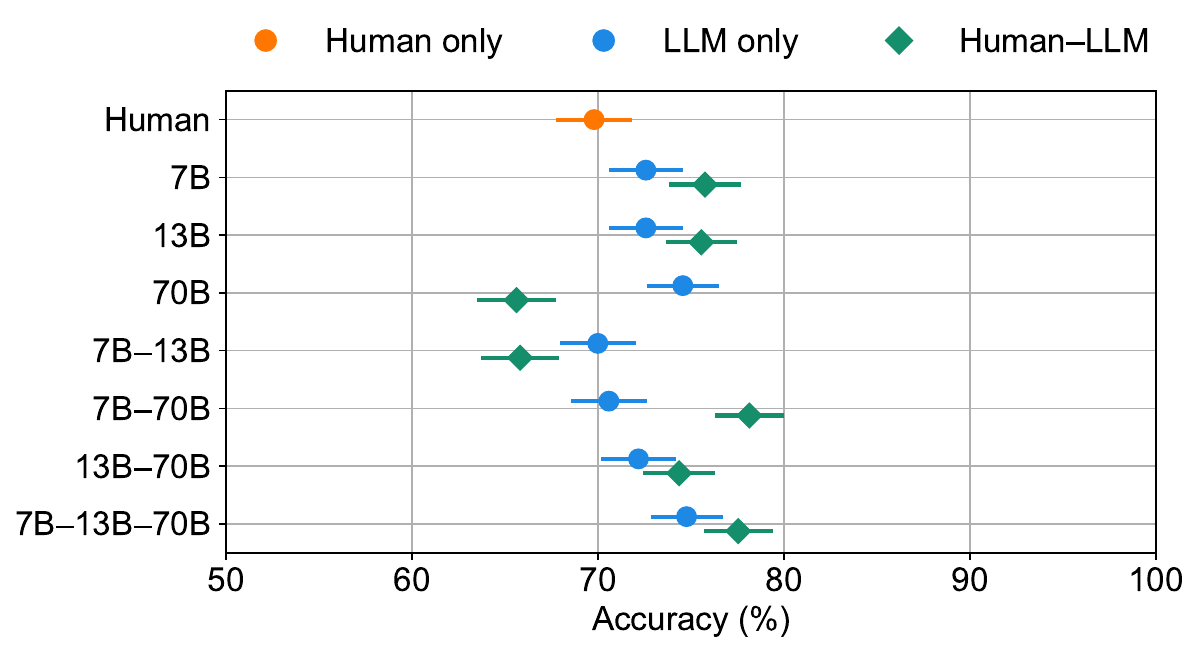}
\caption{\textbf{Removing confidence from the logistic combination model diminishes team performance.}
\newline
Accuracy results on the neuroscience forecasting task 
with the confidence-weighted logistic combination model,
where the magnitude of the confidence scores was set to 1, i.e., 
$f(x)=1$ in Equation~({\ref{eq:squashing}}).
Adding a human to a team with one or more machines (blue points) does not necessarily improve performance (green points). 
Llama2 chat 7B, 13B, and 70B models are considered. 
Each data point corresponds to the average across 503 test case evaluations.
Error bars represent standard error of the mean using a binomial model.}
\label{fig:brainbench_logistic_predictions_no_confidence_experiment}
\end{figure*}

In the object recognition task, 
team performance is not affected by teammates' confidence.
Does the fluctuating confidence on a trial by trial basis matter for LLMs? 
Or could complementarity be achieved by just forming a weighted average 
of responses as in the object recognition task?
We evaluated the impact of confidence by setting the magnitude of 
the confidence scores to 1.
This setting mimics the concept of ``the wisdom of the crowds," 
where responses are considered without factoring in confidence levels.
We find that confidence is particularly important in this knowledge-intensive task, 
as the overall performance diminishes when confidence scores are neglected
(Figure~{\ref{fig:brainbench_logistic_predictions_no_confidence_experiment}}).
With this variant, human--LLM teams do not always surpass LLM-only teams 
(Welch's $t(7.36)=0.38$, $P=0.36$).
The model including confidence (Figure~{\ref{fig:brainbench_logistic_predictions_base_experiment}}) 
outperforms the variant without confidence 
(Figure~{\ref{fig:brainbench_logistic_predictions_no_confidence_experiment}})
by a significant margin (Welch's $t(17.08)=3.16$, $P<0.01$).

We then assessed whether our regression approach would benefit from a
more elaborated formulation.
Equation~({\ref{eq:squashing}}) presents a function that modulates the magnitude
of a team member’s confidence rating to adjust their calibration.
In an optimal setting (Figure~{\ref{fig_supp:squashing}}), the resulting model
was indistinguishable from the standard regression model
(Welch's $t(19.32)=0.05$, $P=0.96$).
Adding interaction terms to the confidence-weighted features does not 
improve the performance of either human--LLM or LLM-only teams
(Figure~{\ref{fig_supp:polynomial}}).
More complex model variants may perform better in other tasks, 
especially when there is more and less noisy data. 
Simpler model variants seem to be more robust to noise in our 
cross-validated experiments.
Thus, the simplicity of our approach provides effective team collaboration.

Our confidence-weighted logistic regression approach follows from 
the principles of a Bayesian combination model that fosters human--machine 
complementarity~\citep{steyvers2022hai} 
(Figure~{\ref{fig_supp:imagenet_80_reproducing_original_study}}).
One questions is how well our logistic regression approach compares to 
the Bayesian approach beyond image classification.
Our confidence-weighted regression model outperformed
(Figure~{\ref{fig:model_comparison}}) the Bayesian model 
when evaluated on the three human--LLM and three LLM--LLM teams
for which the Bayesian model is intended to apply 
(Welch's $t(8.75)=2.91$, $P<0.01$).
This success is impressive given that the regression approach 
takes seconds to compute on a current desktop, whereas 
the Bayesian approach is orders of magnitude slower.
\section{Discussion}
Can humans team effectively with machines such as LLMs when the humans perform worse? 
We developed a confidence-weighted regression approach that can 
integrate judgments from any number of teammates.
Using this method and testing on two forecasting benchmarks~\citep{steyvers2022hai,luo_large_2024},
we found that human--machine teams achieve complementarity, 
that is their combined performance bests that of either teammate alone
(Figures~\ref{fig:imagenet_125_logistic_predictions_base_experiment}
and \ref{fig:brainbench_logistic_predictions_base_experiment}).
Complementarity was achieved because two critical conditions were satisfied, 
namely confidence was well-calibrated and classification diversity held among teammates 
(Figure~\ref{fig:brainbench_results_llama2_chat}).
Strikingly, every combination of machines benefited from adding a human to the team
(Figures~\ref{fig:imagenet_125_logistic_predictions_base_experiment}
and \ref{fig:brainbench_logistic_predictions_base_experiment}).

\begin{figure*}[ht!]
\centering\includegraphics[width=0.8\textwidth]{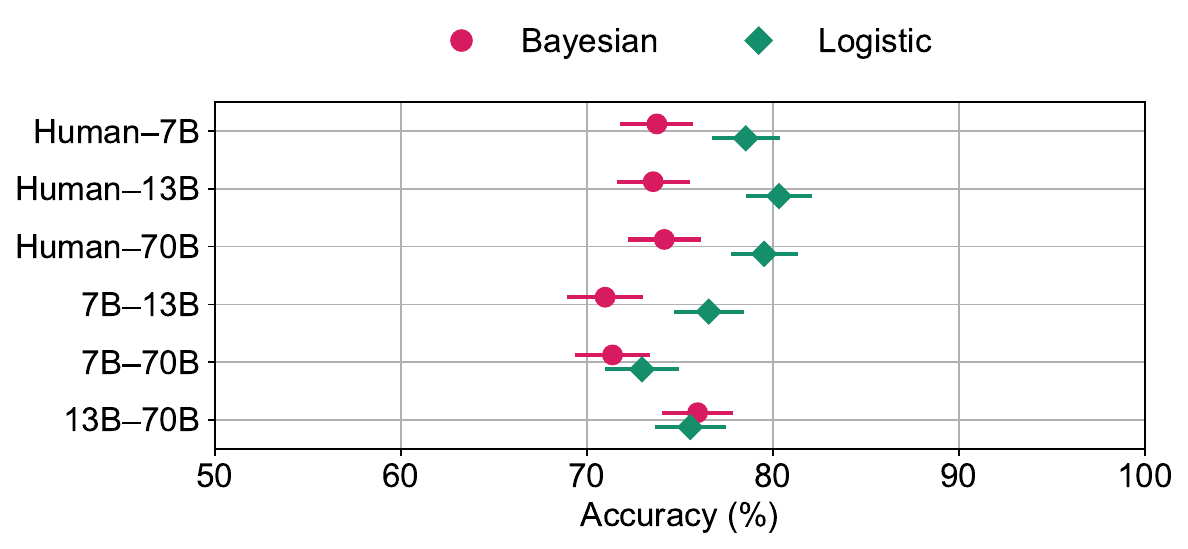}
\caption{\textbf{Comparison between Bayesian and confidence-weighted logistic combination models for human--LLM and LLM--LLM teams.}
\newline
The confidence-weighted logistic combination model more effectively integrates human and machine judgments.
Llama2 chat 7B, 13B, and 70B models are considered.
Each data point corresponds to the average across 503 test case evaluations.  
Error bars represent standard error of the mean using a binomial model.}
\label{fig:model_comparison}
\end{figure*}

Confidence in individual responses significantly impacted team performance 
in the knowledge-intensive task.
The overall performance diminished when confidence scores were neglected 
(Figure~\ref{fig:brainbench_logistic_predictions_no_confidence_experiment}).
However, in the noisy object recognition task, this effect was not present (Figure~\ref{fig_supp:imagenet_125_logistic_predictions_no_confidence_experiment}).
In both scenarios, the performance of the 
confidence-weighted integration of judgments from teammates
was equal to or better than the weighted average of responses, i.e., 
``the wisdom of the crowds." For the confidence-weighted model variants to thrive, confidence judgments need to be calibrated such that higher confidence is associated with higher accuracy.

Our approach was informed by a Bayesian method for combining judgments of humans and machines~\citep{steyvers2022hai}. Our approach has a number of advantages, including  ease of implementation, very fast runtime, an interpretable solution, and readily extendable to any number of teammates. Surprisingly, our confidence-weighted regression approach performed better than  the Bayesian approach (Figure~\ref{fig:model_comparison}). One possibility is that the discretization of continuous confidence measures, which the Bayesian model requires, limited its performance. Perhaps an alternative formulation would perform better. Unfortunately, reformulating the Bayesian model and properly implementing it requires substantial effort and expertise. In contrast, because our confidence-weighted integration model is formulated within a regression framework, it is straightforward to extend the formulation. For example, by including a function that modulates the calibration of a teammate (Figure~\ref{fig_supp:squashing}), or nonlinear relationships (e.g., polynomial terms) between confidence-weighted predictions and outcomes (Figure~\ref{fig_supp:polynomial}).

While we selected three LLMs with superhuman performance on BrainBench, these LLMs are not the highest performing models on this benchmark~\citep{luo_large_2024}. Our choice was deliberate because
a vastly superior teammate may hinder complementarity. In the limit, a teammate who is never wrong does not need to be part of a team.
This limiting condition may become more prevalent should LLMs continue to improve and, therefore, diminish the benefits of human--LLM teaming. 
For the foreseeable future, we suspect there will be tasks for which humans and LLMs can effectively team. Moreover, our method for integrating the judgments of teammates is not limited to human--LLM teams. Instead, the method is general and applies to any set of agents (natural or artificial) that can report how confident they are in their decisions.

This study explored the possibility of a collaborative approach between humans 
and machines for superior decision-making in classifying noisy natural images and
forecasting neuroscience outcomes.
Our confidence-weighted regression method effectively combined human and machine
judgments because teammates fulfilled the conditions of well-calibrated confidence 
and classification diversity.
Our results suggest there is a place for humans in teams with machines even when the machines perform better.
We hope our work facilitates successful collaborations between humans
and machines in addressing important challenges.
\section{Methods}\label{sec:methods}

\subsection*{Datasets}

\subsubsection*{ImageNet 16H~\citep{steyvers2022hai}}
A subset of the 2012 Large Scale Visual Recognition Challenge (LSVRC) ImageNet
training set~\citep{ILSVRC} was utilized.
Namely, the dataset comprised 1\,200 test cases (i.e., images), 
divided equally into 16 classes 
(chair, oven, knife, bottle, keyboard, clock, boat, bicycle, airplane, truck, 
car, elephant, bear, dog, cat, and bird). 
Four levels of phase noise were independently applied to distort the image dataset ($\Omega=\{80,95,110,125\}$). We considered two noise levels: low ($\Omega=80$), where humans outperform machines (Figure~{\ref{fig_supp:imagenet_80_logistic_predictions_base_experiment}}), and high ($\Omega=125$), where most machines outperform humans (Figure~{\ref{fig:imagenet_125_logistic_predictions_base_experiment}}).

One hundred forty-five participants classified between 34 and 74 noisy images
(i.e., test cases) into the 16 aforementioned categories. 
For each evaluation, participants also provided a discrete confidence 
level (low, medium, or high).
The total number of human classifications corresponded to 7\,247 for low noise 
images, and 7\,239 for high noise images.

Five different machine classifiers pretrained for ImageNet 16H were utilized: 
AlexNet~\citep{AlexNet}, 
DenseNet161~\citep{DenseNet}, 
GoogleNet~\citep{GoogleNet}, 
ResNet152~\citep{ResNet}, and 
VGG-19~\citep{VGG}.
One pass through the noisy images data (epoch) was performed
during stochastic gradient training.
For a given image, the classifiers produced probability scores 
for each of the 16 classes.
The class label was assigned to the class with the highest probability. 

\subsubsection*{BrainBench~\citep{luo_large_2024}}
The benchmark includes test cases created either by
expert neuroscientists or by prompting GPT-4 (Azure OpenAI API; version 2023-05-15)
to create test cases.
Since LLMs outperform humans by a large margin in both scenarios~\citep{luo_large_2024} 
and substantial differences in performance may preclude 
complementarity~\citep{steyvers2022hai}, 
we used the GPT-4 generated test cases because the performance difference between humans and LLMs 
was smaller for these test cases, though LLMs were still clearly superior.
We considered a dataset comprising 100 machine-generated test cases.
These test cases were created from abstracts in the 
\emph{Journal of Neuroscience} published in 2023. 
These abstracts are categorized into five sections: Behavioral/Cognitive, 
Systems/Circuits, Development/Plasticity/Repair, Neurobiology of Disease, 
and Cellular/Molecular.
Each test case contains a published abstract and an altered version produced 
by GPT-4. 
These modifications, though minimal, significantly change the 
results\textemdash for instance, by changing the roles of brain regions or 
reversing a result's direction (e.g., from ``decreases'' to ``increases''). 
The altered abstracts remain logically coherent despite the changes.
The BrainBench task is to identify the correct study outcome by choosing 
between the original abstract and its altered counterpart.

One hundred seventy-one neuroscience experts were recruited to complete 
an online study~\citep{luo_large_2024}. 
Each participant evaluated three out of the 100 test cases.
Two versions of an abstract were presented, 
one with the actual results and one that was altered
(Figure~{\ref{fig:brainbench}}).
Participants chose the version they believed to be the original and rated their confidence using 
a slider bar.
After applying several exclusion criteria, the 171 participants yielded 503 observations (2 to 9 instances per test case).

We considered LLMs from the Llama 2 chat family with 7B, 13B, and 70B weights~\citep{touvron2023llama2openfoundation}. LLMs chose the version of the abstract with the lower perplexity (PPL). Confidence was calculated as the absolute difference in PPL between the original and altered versions of the abstract (Figure~{\ref{fig:brainbench}}).

\subsection*{Bayesian combination model}
Human and machine judgments were combined by adapting a Bayesian framework for 
human--machine complementarity~\citep{steyvers2022hai}. 
The problem setting for combining two team members,
human and machine, is as follows: Let $N$ denote the number of test cases 
to be analyzed with $L$ possible choices.
The ground truth labels of the original test cases are $z\in\{0,\ldots,L-1\}^{N}$.
For the human classifier, the predicted labels $y\in\{0,\ldots,L-1\}^{N}$ and 
their corresponding confidence ratings $r\in\{0,\ldots,R-1\}^{N}$ 
with ``0: lowest possible confidence'' and
``$R-1$: highest possible confidence'' are given. 
For the machine classifier, we used the probability scores 
$\pi\in\mathbb{R}_+^{N\times L}$.
In the case of LLMs, $\pi = \text{Softmax}(-q)$, where 
$q$ are the PPL scores, reflecting
a measure of uncertainty~\citep{luo_large_2024}. 

The first step of this model
is to generate correlated probability scores for human and machine 
classifiers using a bivariate normal distribution,
\begin{equation}\label{eq:probscores}
\left(\begin{matrix}\pi_H \\ \pi_M \end{matrix}\right)\sim
\mathcal{N}\left(\left(\begin{matrix} \mu_H \\ \mu_M \end{matrix}\right), 
    \left(\begin{matrix}\sigma_H^2 & \sigma_H\sigma_M\rho_{HM}\\ \sigma_H\sigma_M\rho_{HM} & \sigma_M^2\end{matrix}\right)\right)\,.
\end{equation}
The means of the underlying distribution, $\mu_H$ and $\mu_M$, 
depend whether the label of test case $i$, 
$z_i$, is correct or not, i.e.,
\begin{equation*}
\begin{aligned}
\mu_{i,j,H} &= b_H + (a_H - b_H)\cdot\mathbf{1}_{\mathcal{Z}_i}(j)\,,\\
\mu_{i,j,M} &= b_M + (a_M - b_M)\cdot\mathbf{1}_{\mathcal{Z}_i}(j)\,,
\end{aligned}
\end{equation*}
with $\mathcal{Z}_i=\{x\,|\,x=z_i\}$.
Note that the scalar parameters 
$a_H$, $a_M$, $b_H$, $b_M$, $\sigma_H$, $\sigma_M$, and $\rho_{HM}$ 
in Equation~({\ref{eq:probscores}}) are learned from data.
The parameter $\rho_{HM}$ from the covariance matrix learns
the correlation between the human and machine classifiers.
In the case of the machine classifier, 
$\pi_M$ is compared to the empirical probability scores, $\pi$. 
Then, for the human classifier, $\pi_H$ is a latent variable that is used 
to calculate classifications
\begin{equation*}
y_H \sim \text{Categorical}\left(\text{Softmax}\left(\pi_{H}/\tau\right)\right)\,,
\end{equation*}
where $\tau$ denotes a temperature parameter, usually small, that helps 
convergence~\citep{steyvers2022hai}. The predicted classifications, $y_H$,
are compared to actual human predictions, $y$.
Ordered logit and probit models~\citep{mccullagh1980ologit} yield
practically indistinguishable results~\citep{Boes2006}.
Departing from Steyver's et al.'s~\citep{steyvers2022hai}, 
we used ordered logit because of software availability. 
We successfully reproduced their prior results with probit using ordered logit 
(see Figure~{\ref{fig_supp:imagenet_80_reproducing_original_study}}). 
Ordered logit maps the continuous probability scores, $\pi_H$,
to an ordinal confidence rating, $r_H$.
This means,
\begin{equation}\label{eq:orderedlogistic}
r_H \sim \text{OrderedLogistic}(\pi_{H}, c, \delta)\,,
\end{equation}
where the parameters $c\in\mathbb{R}_+^{R-1}$ are the breakpoints of the intervals 
that map $\pi_{H}$ into $r_H$, and $\delta$ is a scalar that controls the sharpness 
of the rating probability curves.
Finally, $r_H$ is compared to the empirical human confidence ratings, $r$.

\subsection*{Confidence-weighted logistic combination model}
We introduce a logistic regression approach that combines the judgments of any number of teammates. The logistic combination model follows the principles of the Bayesian combination model, but is formulated within an easier to implement and extend  regression framework. In its most basic form, which we consider here, each teammate is captured by a single predictor in the regression model. The value of the predictor on a trial depends on the teammate's choice and their confidence. In particular, the magnitude of the predictor is the teammate's confidence on that trial (i.e., confidence-weighted integration) and the sign is determined by the teammate's choice. In general, the fitted beta weight for a teammate will reflect their accuracy and calibration.

As in the Bayesian combination model, 
$y$, $r$, $\pi$ are given.
The logistic function is of the form:
\begin{equation*}
p_x = \frac{1}{1 + e^{-\beta^\T x}}\,,
\end{equation*}
where $p_x$ is the predicted probability of the arbitrarily assigned first 
option, and the evidence is
\begin{equation}
\beta^\T x = \beta_I + \beta_H x_H + \beta_M x_M\,.
\label{eq:evidence}
\end{equation}
The fitted weights $\beta_I$, $\beta_H$, and $\beta_M$ correspond to 
the intercept, and human and machine teammates, respectively.
The term $x_{i,j,k}$ is the signed confidence of the $i$-th test case for the class label $j$ selected by team member $k$. 
For human participants, $x_{i,j,H}$ is $r_i$ if $y_i=j$ is the selected class label
and $-r_i$ otherwise.
Similarly, for machine team members, $x_{i,j,M} = \pi_{i,j}$ if $y_i=j$ is the selected class label
and $x_{i,j,M} = -\pi_{i,j}$ otherwise.
In the binary case of LLMs, we consider the absolute PPL difference as a measure of confidence.
This means, 
$x_{i,0,M}$ is $|\Delta q_i|$ if $y_i=0$ is the selected class label
and $-|\Delta q_i|$ otherwise.

Consider the case where an agent is accurate, but its confidence is not calibrated.
Then, a slightly more complex version of this logistic regression 
can be formulated by introducing a single-parameter function for each team member.
This function can either pass through the magnitude of a team member's confidence rating 
or squash its magnitude toward 1 for every confidence rating, e.g.,
\begin{equation}
f(x) = 1 + \frac{x-1}{1 + \alpha|x-1|}\,.
\label{eq:squashing}
\end{equation}
That more complex model should work better if an agent is a good predictor of 
outcomes but its confidence is random. 
In that case, $\alpha$ will be high, and the magnitude of the confidence weighting 
will always be $f(x)=1$ in Equation~({\ref{eq:squashing}}).
Thus, the fitted weights for logistic regression would not penalize the agent as 
much in the combination.
For a really well-calibrated agent, $\alpha$ should be close to 0. Thus, 
Equation~({\ref{eq:squashing}}) becomes the identity function, $f(x)=x$.

It turns out that for the two tasks, there was no significant difference when 
including a more complex model optimizing $\alpha$. 
However, the no confidence variant  
(Figure~{\ref{fig:brainbench_logistic_predictions_no_confidence_experiment}}) 
was considered as a foil to our approach which values confidence.

Additionally, this model can be easily expanded by including additional terms to 
Equation~({\ref{eq:evidence}}). 
For example, a third fitted weight could be included for an interaction 
term $x_H x_M$. 
Likewise, polynomial regression could be used to include $x_H^2$ and $x_M^2$, 
and corresponding fitted weights. 

\subsection*{Cross-validation procedure}
Given a total of $M$ observations across $N$ different test cases (with $M>N$),
we performed a leave-one-out cross-validation (LOOCV)
that provides the best bias-variance trade-off for small datasets.
We evaluated the consistency of performance estimates across different 
cross-validation procedures, demonstrating that our results reflect true model 
generalization rather than artifacts of a particular validation strategy
(Figure~{\ref{fig_supp:CV_procedures}}).

Consider the evaluation of the $i$-th test case.
With this procedure, 
we removed all instances of test case $i$ leaving 
the remaining $N-1$ test cases with all their instances
to train the classifier teams.
For testing, we utilized all the instances of test case $i$.
This was repeated for all $N$ test cases, yielding $M$ predictions.
Note that for individual teammates the evaluation comprised only the testing phase. 
When the predicted labels of any team or individual teammate in this study 
were randomly shuffled, the LOOCV accuracy dropped approximately 
to chance level (i.e.,~$1/L$).

\section*{Data and code availability}
For ImageNet 16H~\citep{steyvers2022hai} and 
BrainBench~\citep{luo_large_2024}, 
the previously reported human participant data and machine confidence scores 
utilized in each study are available at
\url{https://osf.io/2ntrf}
and \url{https://github.com/braingpt-lovelab/BrainBench}, respectively.
All computer code associated with this work including combination model 
implementations, team evaluations, and analyses are publicly available at
\url{https://github.com/braingpt-lovelab/haico}
and archived in Edmond~\citep{edmond}.

\section*{Acknowledgments}
This work was supported by 
a NeuroData Discovery Award from the Kavli Foundation to F.Y.,
the ESRC (ES/W007347/1), 
Microsoft (Accelerate Foundation Models Research Program),
a Royal Society Wolfson Fellowship (18302), and
an AI safety grant from the Foresight Institute to B.C.L.

\section*{Author contributions}
Conceptualization, F.Y. and B.C.L; 
methodology, F.Y., X.L., and B.C.L; 
investigation, F.Y. and X.L.; 
writing-–original draft, F.Y.; 
writing-–review \& editing, F.Y., X.L., and B.C.L; 
funding acquisition, F.Y. and B.C.L;  
resources, F.Y. and O.V.M.;  
supervision, B.C.L.


\section*{Declaration of interests}
The authors declare no competing interests.

\bibliography{references}

\begin{thebibliography}{30}
\providecommand{\natexlab}[1]{#1}
\providecommand{\url}[1]{\texttt{#1}}
\expandafter\ifx\csname urlstyle\endcsname\relax
  \providecommand{\doi}[1]{doi: #1}\else
  \providecommand{\doi}{doi: \begingroup \urlstyle{rm}\Url}\fi

\bibitem[Eppler and Mengis(2004)]{eppler2004_overload}
Martin~J. Eppler and Jeanne Mengis.
\newblock {The Concept of Information Overload: A Review of Literature from
  Organization Science, Accounting, Marketing, MIS, and Related Disciplines}.
\newblock \emph{The Information Society}, 20\penalty0 (5):\penalty0 325--344,
  2004.
\newblock \doi{10.1080/01972240490507974}.

\bibitem[Bawden and Robinson(2009)]{bawden2009_overload}
David Bawden and Lyn Robinson.
\newblock {The dark side of information: overload, anxiety and other paradoxes
  and pathologies}.
\newblock \emph{Journal of Information Science}, 35\penalty0 (2):\penalty0
  180--191, 2009.
\newblock \doi{10.1177/0165551508095781}.

\bibitem[LeCun et~al.(2015)LeCun, Bengio, and Hinton]{deeplearning2015}
Yann LeCun, Yoshua Bengio, and Geoffrey Hinton.
\newblock {Deep learning}.
\newblock \emph{Nature}, 521:\penalty0 436--444, 2015.
\newblock \doi{10.1038/nature14539}.

\bibitem[Silver et~al.(2016)Silver, Huang, Maddison, Guez, Sifre, van~den
  Driessche, Schrittwieser, Antonoglou, Panneershelvam, Lanctot, Dieleman,
  Grewe, Nham, Kalchbrenner, Sutskever, Lillicrap, Leach, Kavukcuoglu, Graepel,
  and Hassabis]{alphago}
David Silver, Aja Huang, Chris~J. Maddison, Arthur Guez, Laurent Sifre, George
  van~den Driessche, Julian Schrittwieser, Ioannis Antonoglou, Veda
  Panneershelvam, Marc Lanctot, Sander Dieleman, Dominik Grewe, John Nham, Nal
  Kalchbrenner, Ilya Sutskever, Timothy Lillicrap, Madeleine Leach, Koray
  Kavukcuoglu, Thore Graepel, and Demis Hassabis.
\newblock {Mastering the game of Go with deep neural networks and tree search}.
\newblock \emph{Nature}, 529:\penalty0 484--489, 2016.
\newblock \doi{10.1038/nature16961}.

\bibitem[Jumper et~al.(2021)Jumper, Evans, Pritzel, Green, Figurnov,
  Ronneberger, Tunyasuvunakool, Bates, Žídek, Potapenko, Bridgland, Meyer,
  Kohl, Ballard, Cowie, Romera-Paredes, Nikolov, Jain, Adler, Back, Petersen,
  Reiman, Clancy, Zielinski, Steinegger, Pacholska, Berghammer, Bodenstein,
  Silver, Vinyals, Senior, Kavukcuoglu, Kohli, and Hassabis]{alphafold}
John Jumper, Richard Evans, Alexander Pritzel, Tim Green, Michael Figurnov,
  Olaf Ronneberger, Kathryn Tunyasuvunakool, Russ Bates, Augustin Žídek, Anna
  Potapenko, Alex Bridgland, Clemens Meyer, Simon A.~A. Kohl, Andrew~J.
  Ballard, Andrew Cowie, Bernardino Romera-Paredes, Stanislav Nikolov, Rishub
  Jain, Jonas Adler, Trevor Back, Stig Petersen, David Reiman, Ellen Clancy,
  Michal Zielinski, Martin Steinegger, Michalina Pacholska, Tamas Berghammer,
  Sebastian Bodenstein, David Silver, Oriol Vinyals, Andrew~W. Senior, Koray
  Kavukcuoglu, Pushmeet Kohli, and Demis Hassabis.
\newblock {Highly accurate protein structure prediction with AlphaFold}.
\newblock \emph{Nature}, 596:\penalty0 583--589, 2021.
\newblock \doi{10.1038/s41586-021-03819-2}.

\bibitem[Zhang et~al.(2024)Zhang, Ladhak, Durmus, Liang, McKeown, and
  Hashimoto]{news-summarization}
Tianyi Zhang, Faisal Ladhak, Esin Durmus, Percy Liang, Kathleen McKeown, and
  Tatsunori~B. Hashimoto.
\newblock {Benchmarking Large Language Models for News Summarization}.
\newblock \emph{Transactions of the Association for Computational Linguistics},
  12:\penalty0 39--57, 2024.
\newblock \doi{10.1162/tacl_a_00632}.

\bibitem[Radford et~al.(2019)Radford, Wu, Child, Luan, Amodei, and
  Sutskever]{gpt2}
Alec Radford, Jeff Wu, Rewon Child, David Luan, Dario Amodei, and Ilya
  Sutskever.
\newblock {Language Models are Unsupervised Multitask Learners}.
\newblock \emph{OpenAI blog}, 2019.

\bibitem[Brown et~al.(2020)Brown, Mann, Ryder, Subbiah, Kaplan, Dhariwal,
  Neelakantan, Shyam, Sastry, Askell, Agarwal, Herbert-Voss, Krueger, Henighan,
  Child, Ramesh, Ziegler, Wu, Winter, Hesse, Chen, Sigler, Litwin, Gray, Chess,
  Clark, Berner, McCandlish, Radford, Sutskever, and Amodei]{LLMs_few-shots}
Tom Brown, Benjamin Mann, Nick Ryder, Melanie Subbiah, Jared~D Kaplan, Prafulla
  Dhariwal, Arvind Neelakantan, Pranav Shyam, Girish Sastry, Amanda Askell,
  Sandhini Agarwal, Ariel Herbert-Voss, Gretchen Krueger, Tom Henighan, Rewon
  Child, Aditya Ramesh, Daniel Ziegler, Jeffrey Wu, Clemens Winter, Chris
  Hesse, Mark Chen, Eric Sigler, Mateusz Litwin, Scott Gray, Benjamin Chess,
  Jack Clark, Christopher Berner, Sam McCandlish, Alec Radford, Ilya Sutskever,
  and Dario Amodei.
\newblock {Language Models are Few-Shot Learners}.
\newblock In H.~Larochelle, M.~Ranzato, R.~Hadsell, M.F. Balcan, and H.~Lin,
  editors, \emph{Advances in Neural Information Processing Systems}, volume~33,
  pages 1877--1901, 2020.

\bibitem[Brynjolfsson and McAfee(2014)]{brynjolfsson2014}
Erik Brynjolfsson and Andrew McAfee.
\newblock \emph{{The Second Machine Age: Work, Progress, and Prosperity in a
  Time of Brilliant Technologies}}.
\newblock W. W. Norton \& Company, New York, NY, USA, 2014.

\bibitem[Frey and Osborne(2017)]{FREY2017254}
Carl~Benedikt Frey and Michael~A. Osborne.
\newblock {The future of employment: How susceptible are jobs to
  computerisation?}
\newblock \emph{Technological Forecasting and Social Change}, 114:\penalty0
  254--280, 2017.
\newblock \doi{10.1016/j.techfore.2016.08.019}.

\bibitem[Vaccaro et~al.(2024)Vaccaro, Almaatouq, and Malone]{Vaccaro2024}
Michelle Vaccaro, Abdullah Almaatouq, and Thomas Malone.
\newblock {When combinations of humans and AI are useful: A systematic review
  and meta-analysis}.
\newblock \emph{Nature Human Behaviour}, 8:\penalty0 2293--2303, 2024.
\newblock \doi{10.1038/s41562-024-02024-1}.

\bibitem[Hemmer et~al.(2025)Hemmer, Schemmer, Kühl, Vössing, and
  Satzger]{hemmer2024complementarity}
Patrick Hemmer, Max Schemmer, Niklas Kühl, Michael Vössing, and Gerhard
  Satzger.
\newblock Complementarity in human-ai collaboration: concept, sources, and
  evidence.
\newblock \emph{European Journal of Information Systems}, pages 1--24, 2025.
\newblock \doi{10.1080/0960085X.2025.2475962}.

\bibitem[Steyvers et~al.(2022)Steyvers, Tejeda, Kerrigan, and
  Smyth]{steyvers2022hai}
Mark Steyvers, Heliodoro Tejeda, Gavin Kerrigan, and Padhraic Smyth.
\newblock {Bayesian modeling of human–AI complementarity}.
\newblock \emph{Proceedings of the National Academy of Sciences}, 119\penalty0
  (11):\penalty0 e2111547119, 2022.
\newblock \doi{10.1073/pnas.2111547119}.

\bibitem[Luo et~al.(2025)Luo, Rechardt, Sun, Nejad, Yáñez, Yilmaz, Lee,
  Cohen, Borghesani, Pashkov, Marinazzo, Nicholas, Salatiello, Sucholutsky,
  Minervini, Razavi, Rocca, Yusifov, Okalova, Gu, Ferianc, Khona, Patil, Lee,
  Mata, Myers, Bizley, Musslick, Bilgin, Niso, Ales, Gaebler, Murty,
  Loued-Khenissi, Behler, Hall, Dafflon, Bao, and Love]{luo_large_2024}
Xiaoliang Luo, Akilles Rechardt, Guangzhi Sun, Kevin~K. Nejad, Felipe Yáñez,
  Bati Yilmaz, Kangjoo Lee, Alexandra~O. Cohen, Valentina Borghesani, Anton
  Pashkov, Daniele Marinazzo, Jonathan Nicholas, Alessandro Salatiello, Ilia
  Sucholutsky, Pasquale Minervini, Sepehr Razavi, Roberta Rocca, Elkhan
  Yusifov, Tereza Okalova, Nianlong Gu, Martin Ferianc, Mikail Khona,
  Kaustubh~R. Patil, Pui-Shee Lee, Rui Mata, Nicholas~E. Myers, Jennifer~K
  Bizley, Sebastian Musslick, Isil~Poyraz Bilgin, Guiomar Niso, Justin~M. Ales,
  Michael Gaebler, N~Apurva~Ratan Murty, Leyla Loued-Khenissi, Anna Behler,
  Chloe~M. Hall, Jessica Dafflon, Sherry~Dongqi Bao, and Bradley~C. Love.
\newblock {Large language models surpass human experts in predicting
  neuroscience results}.
\newblock \emph{Nature Human Behaviour}, 9:\penalty0 305--315, 2025.
\newblock \doi{10.1038/s41562-024-02046-9}.

\bibitem[Steyvers et~al.(2025)Steyvers, Tejeda, Kumar, Belem, Karny, Hu, Mayer,
  and Smyth]{steyvers2024calibrationgapmodelhuman}
Mark Steyvers, Heliodoro Tejeda, Aakriti Kumar, Catarina Belem, Sheer Karny,
  Xinyue Hu, Lukas Mayer, and Padhraic Smyth.
\newblock {What large language models know and what people think they know}.
\newblock \emph{Nature Machine Intelligence}, 7:\penalty0 221--231, 2025.
\newblock \doi{10.1038/s42256-024-00976-7}.

\bibitem[Geirhos et~al.(2018)Geirhos, Temme, Rauber, Sch\"{u}tt, Bethge, and
  Wichmann]{NEURIPS2018_0937fb58}
Robert Geirhos, Carlos R.~M. Temme, Jonas Rauber, Heiko~H. Sch\"{u}tt, Matthias
  Bethge, and Felix~A. Wichmann.
\newblock {Generalisation in humans and deep neural networks}.
\newblock In S.~Bengio, H.~Wallach, H.~Larochelle, K.~Grauman, N.~Cesa-Bianchi,
  and R.~Garnett, editors, \emph{Advances in Neural Information Processing
  Systems}, volume~31, pages 1--13, 2018.

\bibitem[Keren(1991)]{KEREN1991217}
Gideon Keren.
\newblock {Calibration and probability judgements: Conceptual and
  methodological issues}.
\newblock \emph{Acta Psychologica}, 77\penalty0 (3):\penalty0 217--273, 1991.
\newblock \doi{10.1016/0001-6918(91)90036-Y}.

\bibitem[Baranski and Petrusic(1994)]{Baranski1994}
Joseph~V. Baranski and William~M. Petrusic.
\newblock {The calibration and resolution of confidence in perceptual
  judgments}.
\newblock \emph{Perception \& Psychophysics}, 55\penalty0 (4):\penalty0
  412--428, 1994.
\newblock \doi{10.3758/BF03205299}.

\bibitem[Tian et~al.(2023)Tian, Mitchell, Zhou, Sharma, Rafailov, Yao, Finn,
  and Manning]{tian-etal-2023-just}
Katherine Tian, Eric Mitchell, Allan Zhou, Archit Sharma, Rafael Rafailov,
  Huaxiu Yao, Chelsea Finn, and Christopher Manning.
\newblock {Just Ask for Calibration: Strategies for Eliciting Calibrated
  Confidence Scores from Language Models Fine-Tuned with Human Feedback}.
\newblock In Houda Bouamor, Juan Pino, and Kalika Bali, editors,
  \emph{Proceedings of the Conference on Empirical Methods in Natural Language
  Processing}, pages 5433--5442, 2023.

\bibitem[Russakovsky et~al.(2015)Russakovsky, Deng, Su, Krause, Satheesh, Ma,
  Huang, Karpathy, Khosla, Bernstein, Berg, and Fei-Fei]{ILSVRC}
Olga Russakovsky, Jia Deng, Hao Su, Jonathan Krause, Sanjeev Satheesh, Sean Ma,
  Zhiheng Huang, Andrej Karpathy, Aditya Khosla, Michael Bernstein,
  Alexander~C. Berg, and Li~Fei-Fei.
\newblock {ImageNet Large Scale Visual Recognition Challenge}.
\newblock \emph{International Journal of Computer Vision}, 115:\penalty0
  211--252, 2015.
\newblock \doi{10.1007/s11263-015-0816-y}.

\bibitem[Krizhevsky et~al.(2012)Krizhevsky, Sutskever, and Hinton]{AlexNet}
Alex Krizhevsky, Ilya Sutskever, and Geoffrey~E Hinton.
\newblock {ImageNet Classification with Deep Convolutional Neural Networks}.
\newblock In F.~Pereira, C.J. Burges, L.~Bottou, and K.Q. Weinberger, editors,
  \emph{Advances in Neural Information Processing Systems}, volume~25, pages
  1--9, 2012.

\bibitem[Huang et~al.(2017)Huang, Liu, van~der Maaten, and
  Weinberger]{DenseNet}
Gao Huang, Zhuang Liu, Laurens van~der Maaten, and Kilian~Q. Weinberger.
\newblock {Densely Connected Convolutional Networks}.
\newblock In \emph{Proceedings of the IEEE Conference on Computer Vision and
  Pattern Recognition (CVPR)}, pages 4700--4708, 2017.
\newblock \doi{10.1109/CVPR.2017.243}.

\bibitem[Szegedy et~al.(2015)Szegedy, Liu, Jia, Sermanet, Reed, Anguelov,
  Erhan, Vanhoucke, and Rabinovich]{GoogleNet}
Christian Szegedy, Wei Liu, Yangqing Jia, Pierre Sermanet, Scott Reed, Dragomir
  Anguelov, Dumitru Erhan, Vincent Vanhoucke, and Andrew Rabinovich.
\newblock {Going Deeper With Convolutions}.
\newblock In \emph{Proceedings of the IEEE Conference on Computer Vision and
  Pattern Recognition (CVPR)}, pages 1--9, 2015.
\newblock \doi{10.1109/CVPR.2015.7298594}.

\bibitem[He et~al.(2016)He, Zhang, Ren, and Sun]{ResNet}
Kaiming He, Xiangyu Zhang, Shaoqing Ren, and Jian Sun.
\newblock {Deep Residual Learning for Image Recognition}.
\newblock In \emph{Proceedings of the IEEE Conference on Computer Vision and
  Pattern Recognition (CVPR)}, pages 770--778, 2016.
\newblock \doi{10.1109/CVPR.2016.90}.

\bibitem[Simonyan and Zisserman(2015)]{VGG}
Karen Simonyan and Andrew Zisserman.
\newblock {Very Deep Convolutional Networks for Large-Scale Image Recognition}.
\newblock \emph{Preprint at arXiv}, 2015.
\newblock \doi{10.48550/arXiv.1409.1556}.

\bibitem[Touvron et~al.(2023)Touvron, Martin, Stone, Albert, Almahairi, Babaei,
  Bashlykov, Batra, Bhargava, Bhosale, Bikel, Blecher, Ferrer, Chen, Cucurull,
  Esiobu, Fernandes, Fu, Fu, Fuller, Gao, Goswami, Goyal, Hartshorn, Hosseini,
  Hou, Inan, Kardas, Kerkez, Khabsa, Kloumann, Korenev, Koura, Lachaux, Lavril,
  Lee, Liskovich, Lu, Mao, Martinet, Mihaylov, Mishra, Molybog, Nie, Poulton,
  Reizenstein, Rungta, Saladi, Schelten, Silva, Smith, Subramanian, Tan, Tang,
  Taylor, Williams, Kuan, Xu, Yan, Zarov, Zhang, Fan, Kambadur, Narang,
  Rodriguez, Stojnic, Edunov, and Scialom]{touvron2023llama2openfoundation}
Hugo Touvron, Louis Martin, Kevin Stone, Peter Albert, Amjad Almahairi, Yasmine
  Babaei, Nikolay Bashlykov, Soumya Batra, Prajjwal Bhargava, Shruti Bhosale,
  Dan Bikel, Lukas Blecher, Cristian~Canton Ferrer, Moya Chen, Guillem
  Cucurull, David Esiobu, Jude Fernandes, Jeremy Fu, Wenyin Fu, Brian Fuller,
  Cynthia Gao, Vedanuj Goswami, Naman Goyal, Anthony Hartshorn, Saghar
  Hosseini, Rui Hou, Hakan Inan, Marcin Kardas, Viktor Kerkez, Madian Khabsa,
  Isabel Kloumann, Artem Korenev, Punit~Singh Koura, Marie-Anne Lachaux,
  Thibaut Lavril, Jenya Lee, Diana Liskovich, Yinghai Lu, Yuning Mao, Xavier
  Martinet, Todor Mihaylov, Pushkar Mishra, Igor Molybog, Yixin Nie, Andrew
  Poulton, Jeremy Reizenstein, Rashi Rungta, Kalyan Saladi, Alan Schelten, Ruan
  Silva, Eric~Michael Smith, Ranjan Subramanian, Xiaoqing~Ellen Tan, Binh Tang,
  Ross Taylor, Adina Williams, Jian~Xiang Kuan, Puxin Xu, Zheng Yan, Iliyan
  Zarov, Yuchen Zhang, Angela Fan, Melanie Kambadur, Sharan Narang, Aurelien
  Rodriguez, Robert Stojnic, Sergey Edunov, and Thomas Scialom.
\newblock {Llama 2: Open Foundation and Fine-Tuned Chat Models}.
\newblock \emph{Preprint at arXiv}, 2023.
\newblock \doi{10.48550/arXiv.2307.09288}.

\bibitem[McCullagh(1980)]{mccullagh1980ologit}
Peter McCullagh.
\newblock {Regression Models for Ordinal Data}.
\newblock \emph{Journal of the Royal Statistical Society. Series B
  (Methodological)}, 42\penalty0 (2):\penalty0 109--142, 1980.
\newblock \doi{10.1111/j.2517-6161.1980.tb01109.x}.

\bibitem[Boes and Winkelmann(2006)]{Boes2006}
Stefan Boes and Rainer Winkelmann.
\newblock {Ordered response models}.
\newblock \emph{Allgemeines Statistisches Archiv}, 90\penalty0 (1):\penalty0
  167--181, 2006.
\newblock \doi{10.1007/s10182-006-0228-y}.

\bibitem[Yáñez et~al.(2025)Yáñez, Luo, Valerio~Minero, and Love]{edmond}
Felipe Yáñez, Xiaoliang Luo, Omar Valerio~Minero, and Bradley~C. Love.
\newblock {Source code for ``Confidence-weighted integration of human and
  machine judgments for superior decision-making''}.
\newblock \emph{Edmond}, 2025.
\newblock \doi{10.17617/3.IGVPQV}.

\bibitem[Bingham et~al.(2019)Bingham, Chen, Jankowiak, Obermeyer, Pradhan,
  Karaletsos, Singh, Szerlip, Horsfall, and Goodman]{bingham2019pyro}
Eli Bingham, Jonathan~P. Chen, Martin Jankowiak, Fritz Obermeyer, Neeraj
  Pradhan, Theofanis Karaletsos, Rohit Singh, Paul Szerlip, Paul Horsfall, and
  Noah~D. Goodman.
\newblock {Pyro: Deep Universal Probabilistic Programming}.
\newblock \emph{Journal of Machine Learning Research}, 20\penalty0
  (28):\penalty0 1--6, 2019.

\end{thebibliography}

\newpage
\section*{Supplementary information}

\setcounter{figure}{0}
\renewcommand\thefigure{S\arabic{figure}}
\renewcommand{\thealgorithm}{S1}

\subsection*{Implementation of Bayesian combination model}
We implemented the Bayesian combination model~\citep{steyvers2022hai} in Python, 
which was originally developed in JAGS.
The model comprises two stages: 
parameter inference (training) and class label prediction (testing).
Algorithm~{\ref{alg:haico_hm}} illustrates the inference procedure for the human--machine case.
During testing, the posterior samples of the parameters learned in training 
are used to evaluate unseen data. 
For each posterior sample, we compute the joint log-likelihood of the machine scores, 
human classifications, and human confidence ratings under each possible class. 
These log-likelihoods are aggregated across samples and normalized via a softmax transformation.
The predicted class label corresponds to the class with the highest probability.

\begin{figure*}[ht!]
\centering\includegraphics[width=0.9\textwidth]{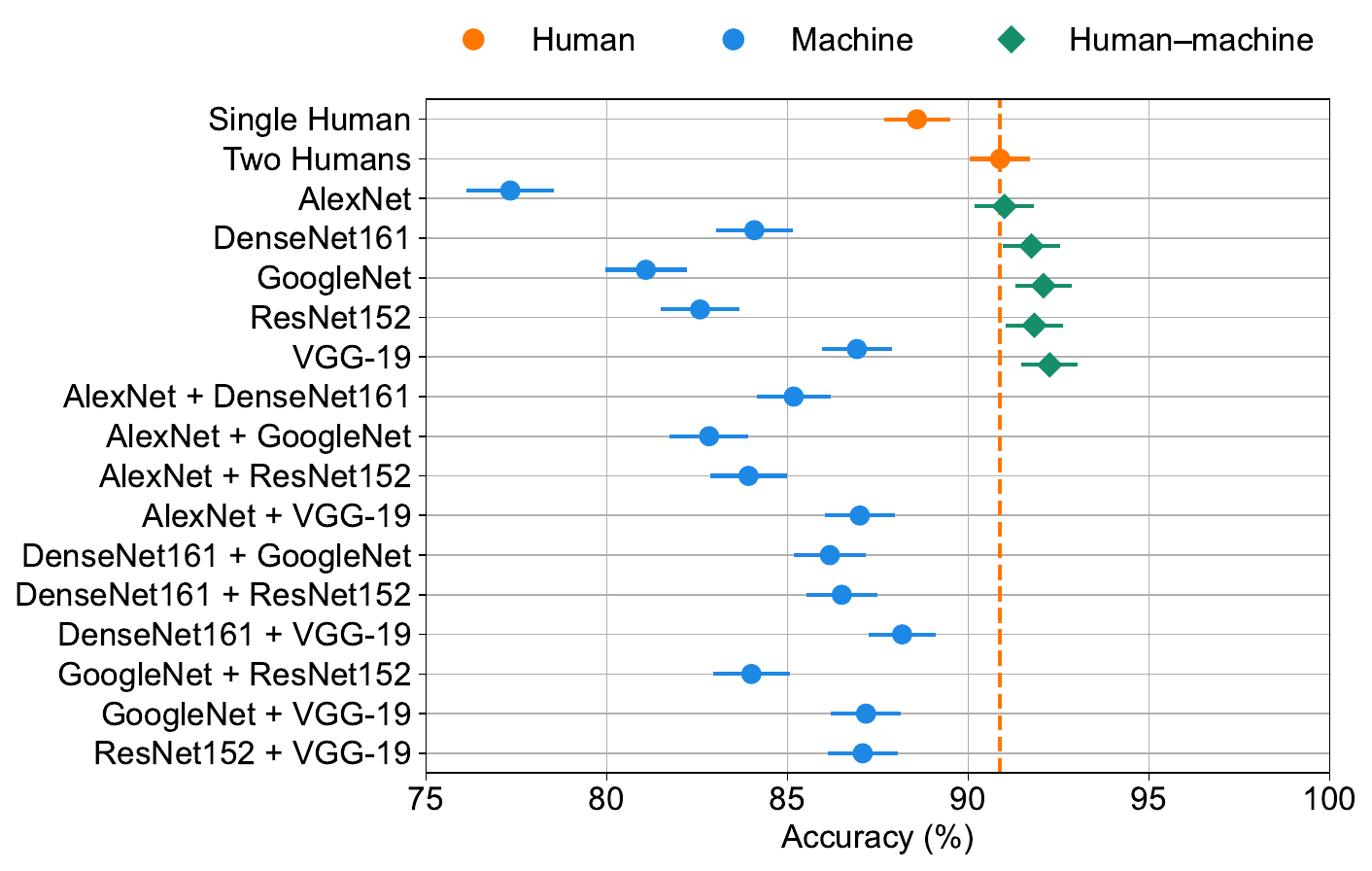}
\caption{\textbf{Performance of the Bayesian combination models in the 
low-noise object recognition task~\citep{steyvers2022hai}.}
\newline
We reproduced Steyver's et al.'s~\citep{steyvers2022hai} results using ordered logit (see Equation~\ref{eq:orderedlogistic}), as opposed to probit.
Accuracy results on low levels of image noise ($\Omega=80$) 
with the Bayesian combination model, reproducing the original setting.
We used 1\,200 predictions (corresponding to 1\,200 unique images) and split them into four random partitions for the purpose of four-fold cross-validation.
Two human participants were created by random sampling without replacement across the 1\,200 unique images. 
In this case, machine classifiers are surpassed by humans ($t(4)=-3.88$, $P<0.01$).
Teams comprising a human and a machine (green points) deliver superior results compared to machine-only and machine--machine teams (Welch's $t(16.04)=9.18$, $P<0.0001$).
Each data point corresponds to the average across 1\,200 image evaluations. 
Error bars represent standard error of the mean using a binomial model.}
\label{fig_supp:imagenet_80_reproducing_original_study}
\end{figure*}

\begin{figure*}[ht!]
\centering\includegraphics[width=0.9\textwidth]{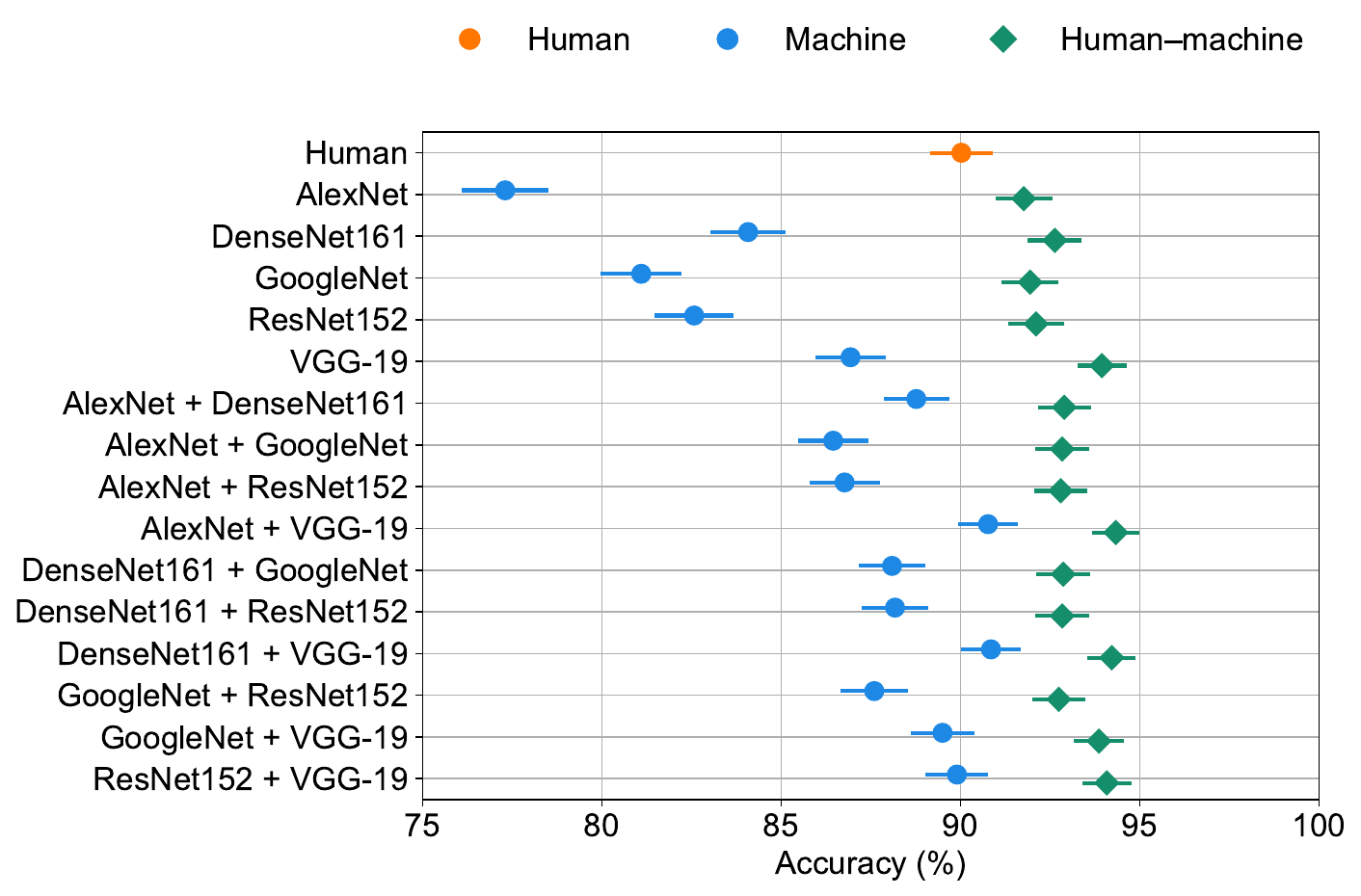}
\caption{\textbf{Performance of the confidence-weighted logistic 
combination model in the low-noise object recognition 
task~\citep{steyvers2022hai}.}
\newline
Accuracy results on low levels of image noise ($\Omega=80$) 
with the confidence-weighted combination model,
where humans outperform machines ($t(4)=-4.77$, $P<0.01$).
Human--machine teams consistently outperform teams with one or more machines (Welch's $t(37.46)=12.50$, $P<0.0001$). 
Each data point corresponds to the average across 7\,247 image evaluations. 
Error bars represent standard error of the mean using a binomial model.}
\label{fig_supp:imagenet_80_logistic_predictions_base_experiment}
\end{figure*}

\begin{figure*}[ht!]
\centering\includegraphics[width=\textwidth]{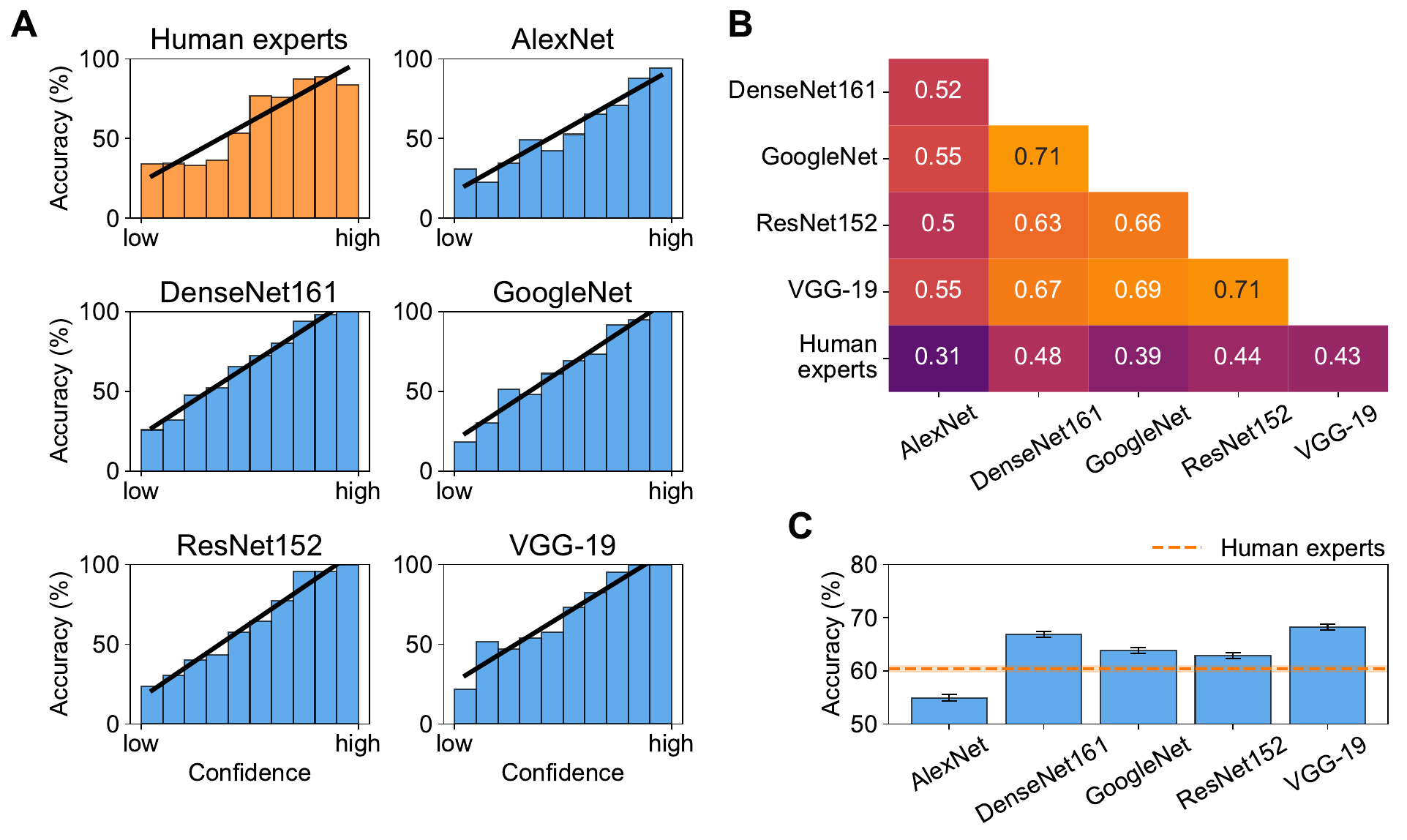}
\caption{\textbf{Conditions for effective collaboration between human experts and machines are satisfied for the noisy object recognition task~\citep{steyvers2022hai}.}
\newline
(A) When human experts and machines were confident in their judgments, they were more likely to be correct. Confidence ratings were sorted into equal bins, and the mean accuracy for each bin was plotted. The positive slope of the black regression lines for humans and models (AlexNet, DenseNet161, GoogleNet, ResNet152, and VGG-19) indicates well-calibrated confidence~\citep{luo_large_2024,KEREN1991217,Baranski1994,tian-etal-2023-just}, meaning higher confidence correlates with higher accuracy.
\newline
(B) Item difficulty Spearman correlations among machines and human experts. For machines, we used the probability score of the predicted image class to determine relative image classification difficulty. The probability score was then signed with $+1$ if the prediction was correct and $-1$ if it was incorrect. Mean accuracy was used for human experts. Heatmap color scale ranges from 0.1 to 0.9.
\newline
(C) Besides AlexNet, models surpass human experts on this task. Error bars represent standard error of the mean using a binomial model.
}
\label{fig_supp:imagenet_calibration}
\end{figure*}

\begin{figure*}[ht!]
\centering\includegraphics[width=0.9\textwidth]{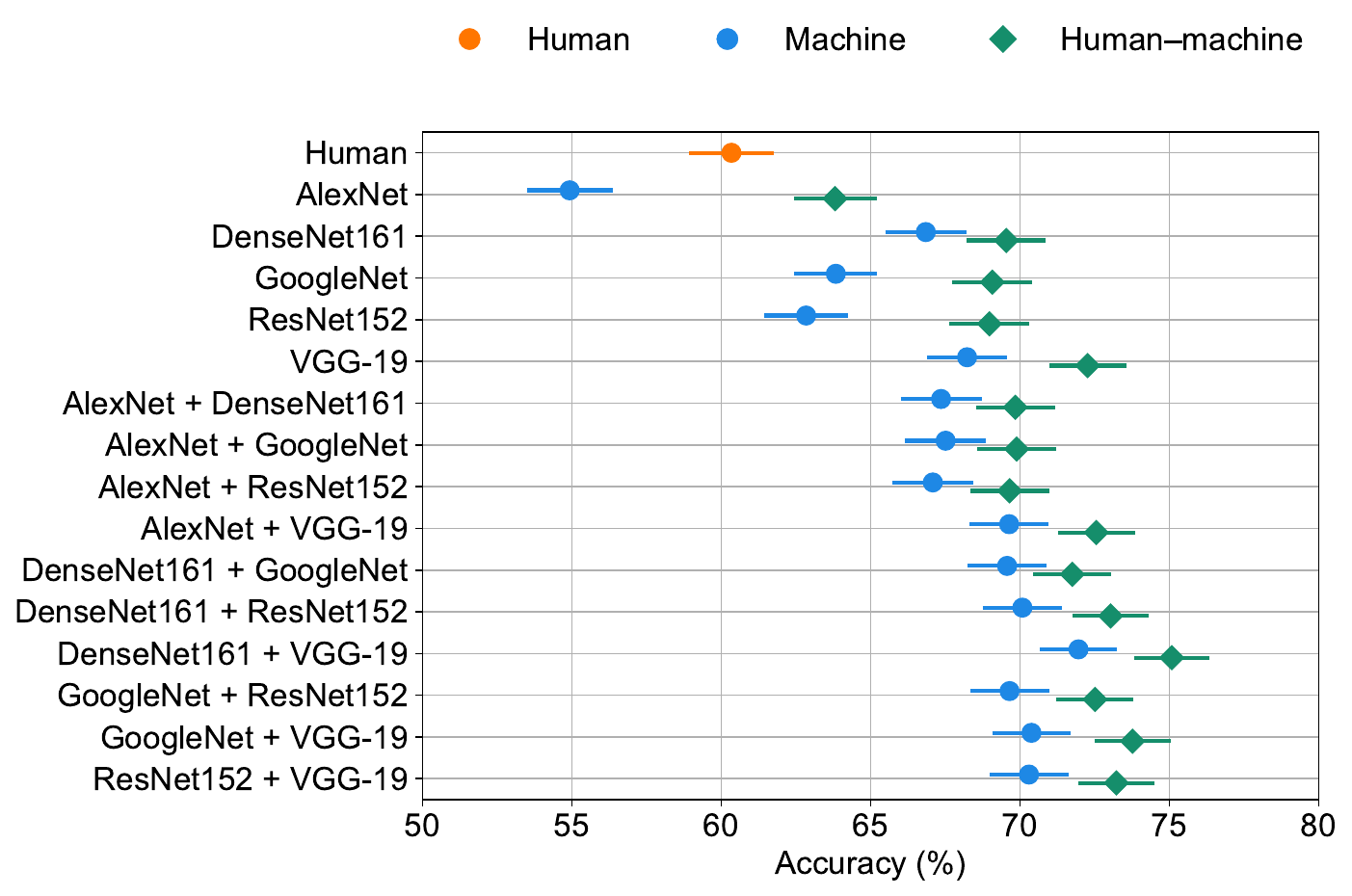}
\caption{\textbf{Removing confidence from the confidence-weighted logistic combination model 
does not negatively impact team performance in the noisy object recognition task~\citep{steyvers2022hai}.}
\newline
Accuracy results on high levels of image noise ($\Omega=125$) 
with the confidence-weighted logistic combination model,
where the probability scores were set to 1, i.e., 
$f(x)=1$ in Equation~\ref{eq:squashing}.
Similarly as in the base scenario (Figure~\ref{fig:imagenet_125_logistic_predictions_base_experiment}), 
human--machine teams surpass machine-only teams (Welch's $t(24.08)=2.80$, $P<0.01$),
although with a smaller margin.
Each data point corresponds to the average across 7\,239 image evaluations. 
Error bars represent standard error of the mean using a binomial model.}
\label{fig_supp:imagenet_125_logistic_predictions_no_confidence_experiment}
\end{figure*}

\begin{figure*}[ht!]
\centering\includegraphics[width=0.9\textwidth]{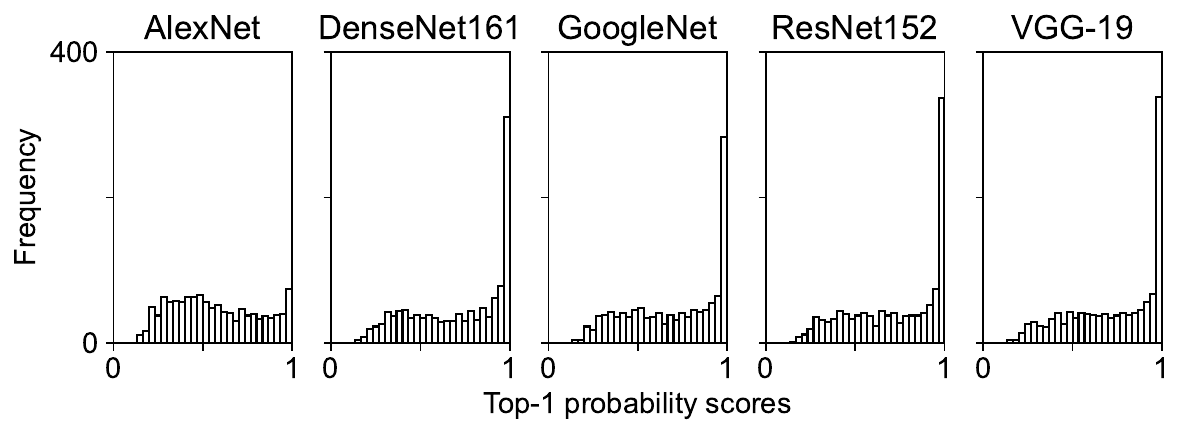}
\caption{\textbf{Confidence assessment in the noisy object recognition task~\citep{steyvers2022hai}.}
\newline
Histogram of top-1 probability scores under high image noise ($\Omega=125$) 
for different machine classifiers. 
For each test case ($n=1200$), only the highest predicted probability was used, 
and values were grouped into 30 bins. 
The resulting distributions are heavily skewed toward 1. 
To test whether confidence magnitudes improve predictions, 
we used cross-entropy, a metric that reflects both accuracy and calibration.
For each classifier, cross-entropy was computed independently for both 
signed probability scores (i.e., signed confidence) and sign-only features. 
The results are nearly identical (Welch's $t(7.99)=0.24$, $P=0.82$), suggesting 
that confidence magnitudes provide no additional benefit in this setting.
}
\label{fig_supp:imagenet_confidence_analysis}
\end{figure*}

\begin{figure*}[ht!]
\centering\includegraphics[width=0.75\textwidth]{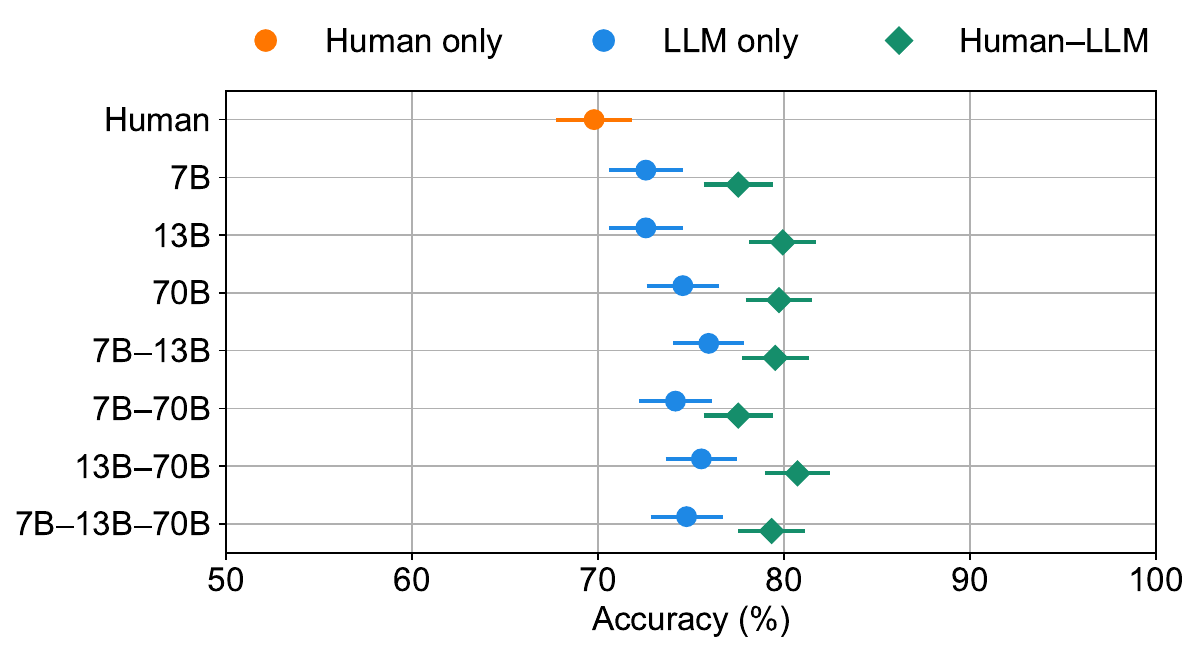}
\caption{\textbf{Optimal squashing in the confidence-weighted logistic combination model does not improve overall team performance in the neuroscience forecasting task~\citep{luo_large_2024}.}
\newline
Accuracy results on the neuroscience forecasting task
with the confidence-weighted logistic combination model,
where the parameter $\alpha$ in Equation~\ref{eq:squashing} was optimized.
Similarly as in the base scenario (Figure~\ref{fig:brainbench_logistic_predictions_base_experiment}),
human--LLM teams surpass LLM-only teams (Welch's $t(11.89)=7.20$, $P<0.0001$).
The performance between base and squashing scenarios is, however, indistinguishable
(Welch's $t(19.32)=0.05$, $P=0.96$).
Each data point corresponds to the average across 503 test case evaluations.
Error bars represent standard error of the mean using a binomial model.}
\label{fig_supp:squashing}
\end{figure*}

\begin{figure*}[ht!]
\centering\includegraphics[width=0.9\textwidth]{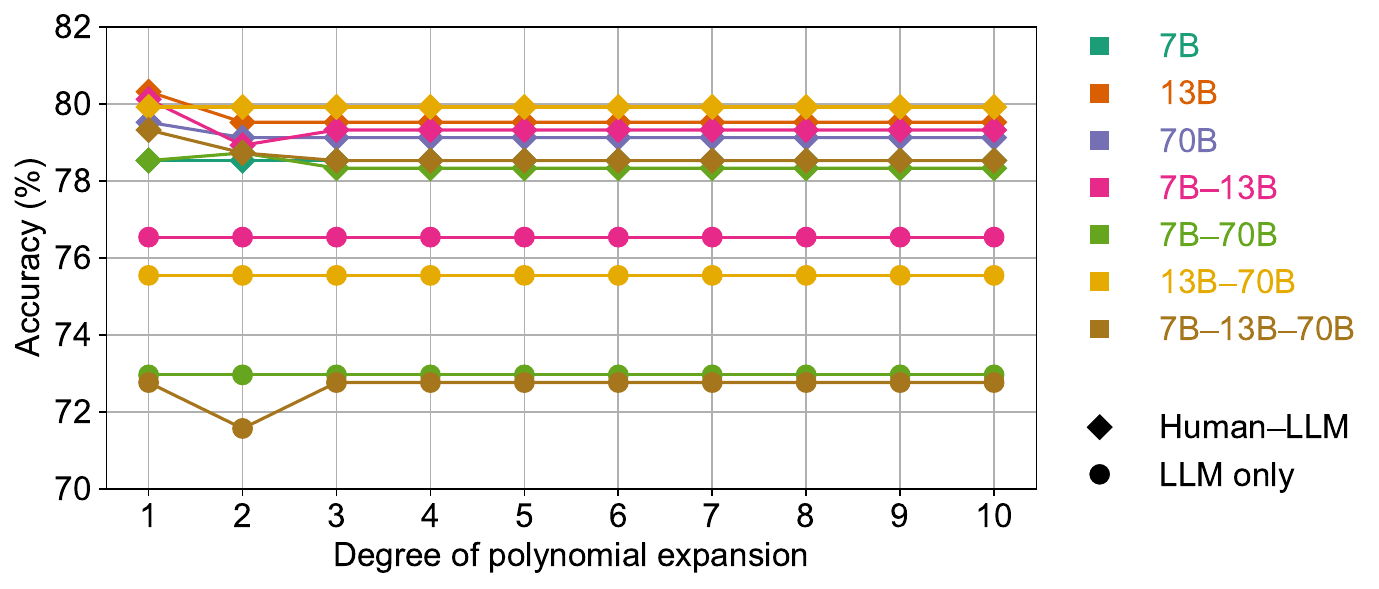}
\caption{\textbf{Adding interaction terms to the confidence-weighted logistic combination model does not improve overall team performance in the neuroscience forecasting task~\citep{luo_large_2024}.}
\newline
Accuracy results of the confidence-weighted logistic combination model 
on the neuroscience forecasting task
as a function of the degree of the polynomial expansion of the features.
Only interaction terms were considered in the polynomial expansion, 
as the results including pure terms were identical.
Team performance is independent of interaction terms.
Each data point corresponds to the average across 503 test case evaluations.}
\label{fig_supp:polynomial}
\end{figure*}

\begin{figure*}[ht!]
\centering\includegraphics[width=0.75\textwidth]{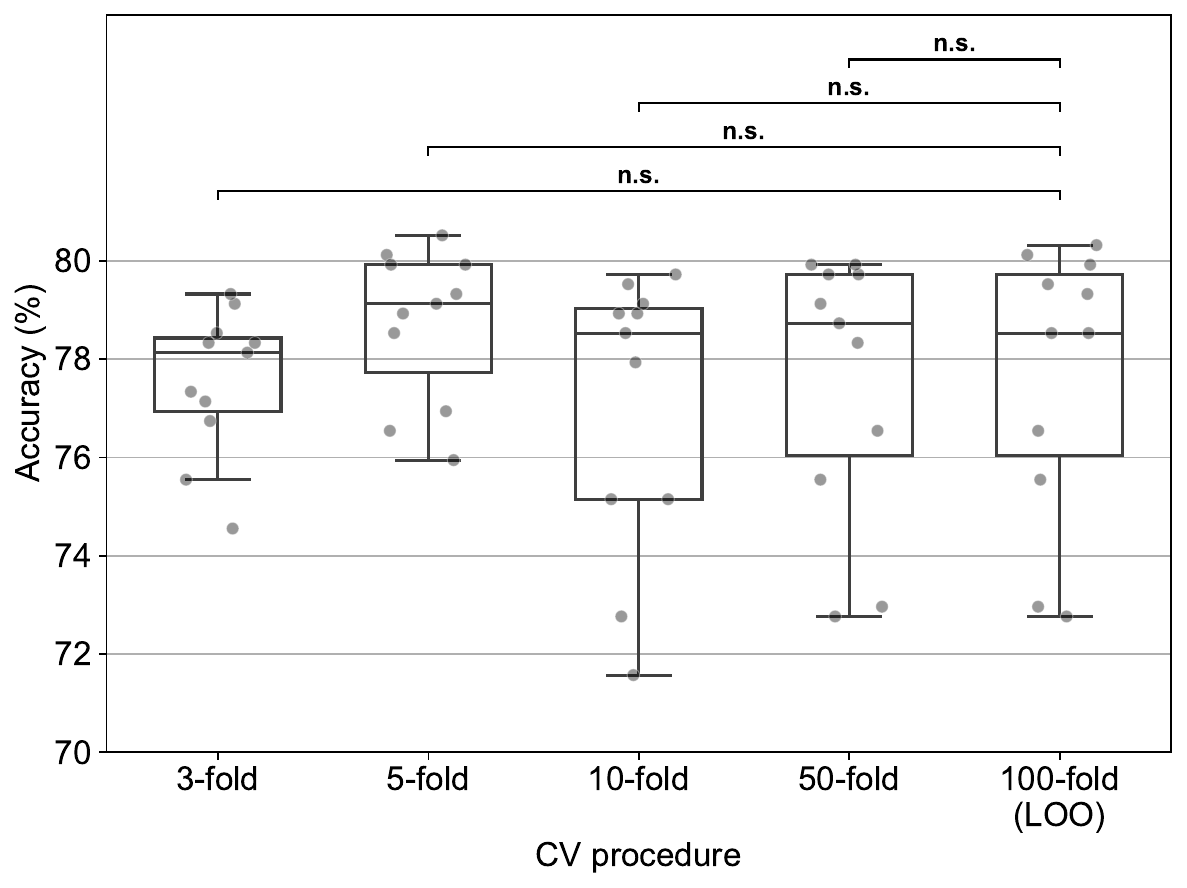}
\caption{\textbf{Performance consistency analysis across different cross-validation procedures.}
\newline
Box plots show the distribution of overall accuracy for 3-fold, 5-fold, 10-fold, 50-fold, 
and leave-one-out (LOO) cross-validation (CV) procedures for the experiment presented in Figure~\ref{fig:brainbench_logistic_predictions_base_experiment}.
Overlaid data points represent human--machine and machine--machine team predictions. 
Paired t-tests (parametric) and Wilcoxon signed-rank tests (non-parametric) were used
to compare each CV procedure with LOOCV to ensure robust assessment regardless of data 
distribution assumptions.
Results show non-significant differences, 
with similar mean accuracies (ranging from 77.0\% to 78.7\%) and 
overlapping 95\% bootstrap confidence intervals (widths between 1.8\% and 3.2\%). 
This demonstrates that LOOCV is an appropriate validation procedure to evaluate model performance.}
\label{fig_supp:CV_procedures}
\end{figure*}

In Figure~{\ref{fig_supp:imagenet_80_reproducing_original_study}}, we reproduced the
experiment from Figure 3 (Top) in Steyver's et al.~\citep{steyvers2022hai} using our Python implementation.
The overall team performance ($n=16$) obtained with our method is indistinguishable from 
the original JAGS implementation (Welch's $t(28.95)=-0.24$, $P=0.81$).
This also holds for the five human--machine (Welch's $t(5.33)=-2.17$, $P=0.08$), 
and ten machine--machine (Welch's $t(17.60)=0.17$, $P=0.87$) teams.

\vspace{3mm}
\algblock{Training}{EndTraining}
\algnewcommand\algorithmictraining{\textbf{start parameter inference:}}
\algnewcommand\algorithmicendtraining{\textbf{end parameter inference}}
\algrenewtext{Training}[1]{\algorithmictraining\ #1}
\algrenewtext{EndTraining}{\algorithmicendtraining}

\begin{algorithm}[ht!]
\caption{Bayesian Combination Model}\label{alg:haico_hm}
\begin{algorithmic}[1]
\vspace{1ex}\State Given data: true class labels $z$, probability scores $\pi_M$, human classification $y$, and confidence ratings $r$.
\vspace{1ex}\State Set priors: $a_M\sim\mathcal{N}(0,10)$, $b_M\sim\mathcal{N}(0,10)$, $\sigma_M\sim\text{Uniform}(0,15)$, 
$a_H\sim\mathcal{N}(0,10)$, $b_H=0$, $\sigma_H=1$, $\rho\sim\text{Uniform}(-1,1)$, $\tau=0.05$, 
$c\sim\text{Uniform}(0,1)$ with $c_i < c_{i+1}\;\forall\,i = 1,\ldots, R-2$, and $\delta\sim\text{Uniform}(0,100)$.
\vspace{3ex}\Training 
\vspace{1ex}\State $\mu_{i,j,M} \gets b_M + (a_M - b_M)\cdot\mathbf{1}_{\mathcal{Z}_i}(j)$
\vspace{1ex}\State $\mu_{i,j,H} \gets b_H + (a_H - b_H)\cdot\mathbf{1}_{\mathcal{Z}_i}(j)$
\vspace{1ex}\State $\pi_{M} \sim\mathcal{N}\left(\mu_{M},\,\sigma_M\right)$ \Comment{Compare to actual data}
\vspace{1ex}\State $\pi_{H} \sim\mathcal{N}\left(\mu_{H} + \rho\,\sigma_H\left(\frac{\pi_{M} - \mu_{M}}{\sigma_M}\right),\,\sqrt{1 - \rho^2}\,\sigma_H\right)$
\vspace{1ex}\State $y \sim \text{Categorical}\left(\text{Softmax}\left(\pi_{H}/\tau\right)\right)$ \Comment{Compare to actual data}
\vspace{1ex}\State $r \sim \text{OrderedLogistic}(\pi_{H}, c, \delta)$ \Comment{Compare to actual data}
\vspace{1ex}\EndTraining
\vspace{1ex}
\end{algorithmic}
\end{algorithm}

We assumed that all human participants shared the same set of parameters 
($a_H$ , $b_H$ , $\sigma_H$, $c$, $\delta$,  and $\tau$).
In the neuroscience forecasting task,
human confidence ratings on the slider bar were mapped to range between 1 and 100.
A wide range is computationally expensive, thus,
we aggregate it into three levels: 
``0: low confidence'', ``1: moderate confidence'', and ``2: high confidence''. 
Then, the aggregated
confidence rating used for analysis, $r\in\{0,1,2\}^{N}$, reads
\begin{equation*}
r =
\begin{cases}
    0 & \text{if }\;\phantom{33 < {}} \text{self-reported confidence} \leq 33, \\
    1 & \text{if }\;33 < \text{self-reported confidence} \leq 66, \\
    2 & \text{if }\;66 < \text{self-reported confidence} \\
\end{cases}\quad.
\end{equation*}
The utilized cutpoints (i.e., 33 and 66) produced a good 
agreement between confidence and accuracy. 
Among the evaluations of human participants, 
``low'' had $63.2\%$ average accuracy ($n=174$), 
``moderate'' had $66.5\%$ ($n=185$), and 
``high'' had $81.9\%$ ($n=144$). 
To infer the posterior over the underlying parameters, 
a No-U-Turn Sampler (NUTS) 
for Markov chain Monte Carlo (MCMC)~\citep{bingham2019pyro} was used with 
$n_w=1000$ warmup steps, $n_c=8$ chains, and $n_s=50$ samples.

\newpage

\newpage

\newpage

\newpage

\newpage

\newpage

\newpage

\newpage


\end{document}